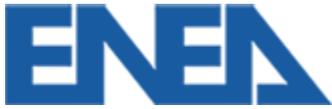

# Swarm robotics and complex behaviour of continuum material

Ramiro dell'Erba
*ENEA Technical Unit technologies for energy and industry – Robotics Laboratory*

**Abstract**
In swarm robotics, just as for an animal swarm in Nature, one of the aims is to reach and maintain a desired configuration. One of the possibilities for the team, to reach this aim, is to see what its neighbours are doing. This approach generates a rules system governing the movement of the single robot just by reference to neighbour's motion. The same approach is used in position based dynamics to simulate behaviour of complex continuum materials under deformation. Therefore, in some previous works, we have considered a two-dimensional lattice of particles and calculated its time evolution by using a rules system derived from our experience in swarm robotics. The new position of a particle, like the element of a swarm, is determined by the spatial position of the other particles. No dynamic is considered, but it can be thought as being hidden in the behaviour rules. This method has given good results in some simple situations reproducing the behaviour of deformable bodies under imposed strain. In this paper we try to stress our model to highlight its limits and how they can be improved. Some other, more complex, examples are computed and discussed. Shear test, different lattice, different fracture mechanism and ASTM shape sample behaviour have been investigated by the software tool we have developed.

## 1. Introduction

In this paper we shall describe the time evolution of a material particles system by a position based dynamics (PBD) method we have developed in previous works (see [1], [2] as references). The advantages of these kinds of methods, with respect to classical finite element methods (FEM) analysis are the following: they do not need to solve computational heavy differential equations and can be easily be used to describe complex object. Moreover they can make use of powerful Graphic Processing Units (GPU) and the task can be parallelized. PBD has been widely used in computer animation due to its efficiency, robustness and simplicity. The aim of the PBD is not to compute physical process but to sacrifice some accuracy to generate visually plausible simulation results with low computational cost [3]. The credibility requirements of user interfaces in videogames has generated a technology, User Interface Physic, where physical principles are partially enforced through ad hoc heuristics assumptions and are often implemented without the usual calculus. The PBD methods result in a physically plausible behaviour of the continuum but suffer from limitations, modelling complex material properties and describing interactions between heterogeneous bodies [4]. This approach has many practical applications. For example a touch screen phone contact list can be scrolled, by fingers, with a motion based on velocity and list length. Reaching the end of the list, the motion will bounce as if a collision occurred. The user feels such behaviour very realistic even if the effects are heuristically reproduced and are not a solution of Newton's law. Other disadvantages of PBD include low fidelity, poor adaptability, and low interactivity. Therefore, sometimes, a physics engine, working through integration techniques that are based on Newton's laws of motion, is added.

Robot swarms are by now a well-known challenge to the scientific community, mainly as an algorithmic tool for the description of collective behaviours, labour division and information sharing [10-13]. Our approach was born indeed by studying flocking rules governing the behaviour of single elements in underwater robotic swarms [5], [6] with the aim of calculating the geometric configuration of a submarine swarm robots by its single elements; this task is very important because the swarm, like school fish, adapt its configuration depending on the mission assigned. To this end, a positioning and control algorithm has been developed so that it reaches the desired configuration. It was then noted that a quite similar algorithm can be adapted to describing PBD problems, because the resulting motion was quite similar to body deformation in some circumstances. Therefore introducing constraints, to be related with constitutive equations of the material, into the relationships describing relative positions between the members of the swarm, we tried to describe the deformation of a Continuum medium. The result is a numerical tool useful to describe complex micro-structures not easily analyzed by Cauchy Continuum theory generating big quantity of experimental data. The proposed approach, as already mentioned, presents several advantages (as well as some limitations) over classical Finite Element (FE) method (see [16-19] for exemplifying interesting applications and [20-23] for cases in which generalized continua are studied by means of FEs), as we will see in the following sections.

In the case of a highly non-homogeneous microstructure it is indeed well known that Classical Cauchy continua are not able to produce accurate predictions, and various kinds of generalizations have to be introduced, either considering additional degrees of freedom to account for the kinematics at the level of the microstructure [24-30] or including in the deformation energy density higher gradients of the displacement than the first one [37-46]. The latter is a particularly relevant topic considering the technological interest in developing exotic mechanical metamaterials able to perform targeted tasks [34-36], and therefore the investigation of new and efficient algorithms is of great interest at the moment.

The model we are proposing can exhibit a rich range of behaviours just changing lattice type and its internal parameters. In this paper we extend the preceding work to some more complex case, such as shear test, different fracture mechanism and ASTM shape sample used in tensile test of materials. The reasons to use an ASTM shape are two; first we would like to investigate the sample behaviour in a more complex case with respect to simple shape. Moreover the use of such a shape has the advantage that it can be compared with a well studied situation.

## 2. The algorithm: a quick resume

The algorithm, realized by Mathematica of the Wolfram Research, to calculate deformation is based on the following assumptions. The two dimensional continuum body is discretized into a finite number of particles occupying, in their initial configuration, the nodes of a lattice. The kind of lattice is chosen between the five plane Bravais lattices; changing lattice we can obtain different results.

We have used also a honey comb lattice that is not a Bravais lattice; for them in fact, all lattice sites are equivalent and any vectors connecting to lattice sites are lattice vectors. These conditions are not satisfied for honeycomb lattice so it is not a Bravais lattice. However the honey comb lattice is important owing to its application in grapheme, so we decided to use it to investigate its behaviour in our tool.

The object is therefore discretized by the chosen lattice. Four kinds of particles are considered, but the modular algorithm is opened to introduce a new kind if it is needed to describe other behaviour; moreover the membership category can be changed with time during body deformation. The leaders represent the first kind, whose motions are assigned, i.e. the imposed strain on part of the body. Their motion is known and determines the motion of the other particles. The followers are the second kind, whose motions are calculated by rules involving the motion of other particles and the characteristics of the lattice. The motion of the followers depends on the position of a certain numbers of neighbours. This results in a constrained geometrical problem leading to a transformation operator between the matrices representing the particles configuration, $C_t$, for a discrete set of time steps $t_1$, $t_2$, ...$t_n$.... the numbers of neighbours to consider for calculus can be varied to obtain different results; for example we can consider the coordination number of the chosen lattice. Particles belonging to the frame are the third kind. To avoid edge effects, like corners collapse, we surround the body by an external frame of point; a shell, so that any follower (including the one on the real boundary) interacts with the same number of elements like the others. The motion of the frame is simple: it only follows the motion of an assigned follower of its competence; in case the assigned followers are more (i.e. in a corner) then an average displacement (or a more generic complex rule) is computed. The frame can be something more complex than a single shell; for example if we are considering second gradient interaction we need a double shell to reach our aim that is the homogeneity of the boundary conditions for all the followers. The fictitious are the last kind of particles. They are ghost-like points introduced in some particular case, for managing fracture. We assume the interactions are decreasing with increasing distance between particles; therefore, when Euclidean distance between points is "great", they lose their interaction. To address the problem we start simply by considering a threshold effect between neighbour elements, so that when the distance overcomes the threshold these elements are no longer taken into account in the calculation of the follower position. To preserve symmetry of the Lagrangian neighbours we introduce the ghost-like points called fictitious elements. They have the purpose of balancing the calculations of the point's displacements. All the properties of these fictitious elements are the same as the followers but their motion is not considered, because their work is just to balance the equations. Where are these ghost elements positioned? Our choice is to put them in a position able to recover the original shape of the lattice. As we have seen [7],[14-15] a change in their assigned position produces effects such as the contraction or loosening of the lattice in the deformed configuration. In fact, varying the distances of the fictitious elements after fracture from the true elements, plastic-like and elastic-like behaviours can be obtained. By elastic behaviour in fracture we mean the property of the fracture edges or of the disconnected pieces originated after fracture has occurred, to recover the original shape. Practically, as can be seen in the flow chart (see Figure1), computing the follower's position there is a check routine on the distances between the follower under examination and its neighbours influencing its new position according to the rules. If the distance between the follower and one of these points is larger than the threshold the point is substituted by a fictitious element. The algorithm can be easily generalized to second gradient by introducing two different thresholds for the two shells of neighbours. In Figure1 a flow chart of the process is shown.

The process is the following. We choose a two dimensional body. Choose one of the Bravais lattice and discretize the body to obtain a discrete matrix to represent it. We now decide the constraint of the lattice and the interaction rules between the followers, in order to describe the correct behaviour of the constitutive equations of the materials. As an example we can decide that the lattice has no constraint and displacement of a follower point is the average value of the displacements of its first neighbours (first gradient). We build an adequate frame to avoid edge effects. We decide the motions of some points, called leaders, for the entire time window we are investigating; we can also decide that they will be leader only for a certain time and late become followers (Category change).

Now we can calculate, for each time step, the new configuration of the lattice in three separate operations. When time increases from $t_0$ to $t_1$ the leaders change their position from initial configuration according to the prescribed equation. So far we have built a new intermediate lattice where only the leaders have been moved. Now we take care that the followers are no longer in equilibrium position owing to the leader's displacement. How we can calculate it? As an example if the interactions rule establishes that a follower has to be in the barycentre of all its neighbours we calculate the new position of each follower, taking into account the leader displacement. So far note that at this stage only the leader's neighbours are involved. Finally we take into account the rules governing the frame displacement. This is our new configuration at time $t_1$. It is important to note that the configuration achieved is not an equilibrium one, because the three operations must be repeated for many time steps, after the leaders stop. To be more clear if at time step $t_1$ the leaders have moved we calculate the follower's displacement. This operation involves only the neighbours of the leaders and not the other far followers. Later we calculate the frame displacement to close the loop. Now there are some followers (the neighbours of the leader's neighbours) that there are no more in equilibrium because there has been the displacement of the leader's neighbours. So we need another time step to adjust the configuration and so on. At a certain time all the followers are involved in the calculation. The followers will suffer the leader's motion after (k-1) time steps where k is the distance from the leaders, measured in layers. In this meaning the leader motion "propagates" through the lattice to influence the position of all the followers in a time depending on the lattice dimensions and how much shell of points being considered in the neighbours definition. In the same way when leaders stop the followers continue to adjust their position in many time steps. We have often used the rule of centre of gravity to determine follower's position that mean:

$$x_j(t) = \frac{\sum_{k=1}^{all\ neighbours\ of\ j} x_k(t)}{N}$$

Where N is the total number of neighbours. The same equation is used for the y coordinate.

But we can use different rules in order to approximate different constitutive equations, i.e. we can introduce relative distance between the points into the rule to weight their influence on the follower's movement and simulate Hook law, where force is increased with increasing deformation:

$$x_j(t) = \frac{\sum_{k=1}^{all\ neighbours\ of\ j} dis(k,j) x_k(t)}{\sum_{k=1}^{all\ neighbours\ of\ j} dis(k,j)}$$

Where dis(k,j) is the Euclidean distance between the points k and j.
Or we can mix x-y coordinates into the rules to make the movement in x direction have effect on the y coordinate (lateral contraction).

$$y_j(t) = K * (x_j(t) - x_j(t_0)) * da + \frac{\sum_{k=1}^{all\ neighbours\ of\ j} y_k(t)}{N}$$

Where da is a function of the distance from the central axis, K a parameter determining the response force and $x(t_0)$ the initial x coordinate. This rule leads to a Poisson effect, because an expansion of x coordinate has influence on the y coordinate.
Moreover we can force the follower's movement to overcome the barycentre equilibrium position leading the lattice to oscillate. This will be done in a future paper.

Practically we have a transformation operator between matrices representing initial and final configuration. Remember that, in the modular algorithm, the neighbours can dynamically change at every time step. The choice to fix the neighbours of every particle at the initial time $t_0$, and not to change them during time evolution of the configurations lies in the desire to imitate a crystalline lattice and therefore to deal with solid phase materials. This means the concept of neighbours is Lagrangian, and neighbourhood is preserved during the time evolution of the system; the only exceptions arising with the fracture algorithm. Also the definition of neighbours is customizable by changing metric; for example we can consider points whose Euclidean distance (weighted or not is another possibility to take into account anisotropies) is less than a threshold or, more physically, the coordination number of the lattice chosen. In case of second gradient we enlarge the set of points with a supplementary shell.

Envisaging the possibility to frame the proposed model in a fully variational setting, which is by no means trivial but would provide clear methodological advantages (see [47] for an introduction and [48-52] for illustrative cases concerning continua with non-classical properties), we also like to introduce pseudoenergetic considerations. In the elastic case we can consider the square of the distance between the actual configuration $C$ and the reference configuration $C_0$. It must be underlined that this artifice has no direct connection with the usual energy definition (this is the reason we use the term pseudoenergy) but could be useful to understand deformation. Therefore we introduce two formulations *PE1* and *PE2* for this concept. The first is given by the value, for each time step and in each point describing the configuration, of the sum, extended to the neighbours, of squares of the differences between the distances of the point from its neighbours minus the distance in the initial configuration i.e.

$$PE1(t,j) = \sum_{k=1}^{all\ neighbours\ of\ j} (dis(t,k,j) - dis(t_0 k,j))^2$$

Where *dis(t,k,j)* is the Euclidean distance between points *k* and *j* at time *t*. This is the formula for the point *j* at time *t*
The reason for this choice lies in the attempt to simulate potential energy of material point subject to Hook law.
To compare time contiguous configuration $C_t$ and $C_{t-1}$ we define for each point *j* and each time *t*

$$PE2(t,j) = ||C_t - C_{t-1}||$$

Where || is the Norm of the vector defined by the point *j* at time *t* and *t-1*.

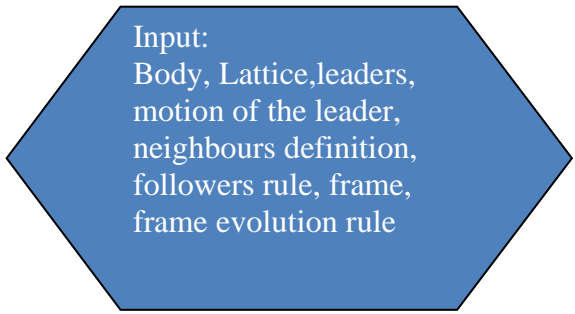

Example of assignment:
1) Body: square
2) Lattice: square
3) Leaders: first line on the right
4) Leaders motion: constant speed for ten time step, later stop
5) Neighbours definition: Nearest coordination number
6) Follower rule: its coordinate are the barycentre of its neighbours
7) Frame: simple
8) Frame rule: each point has same movement of the assigned point

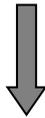
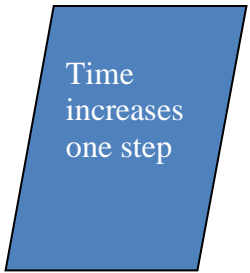
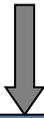
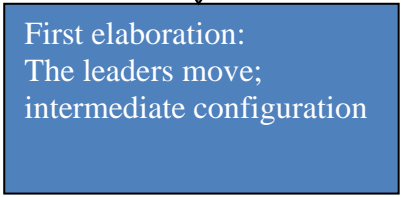

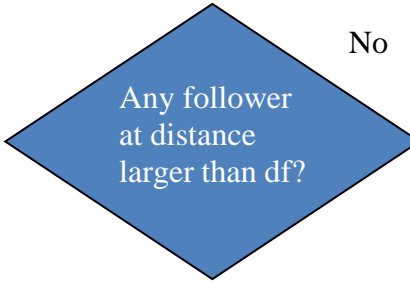

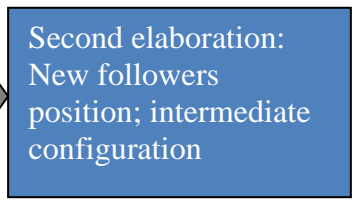

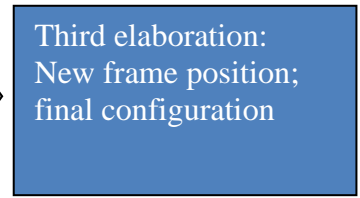
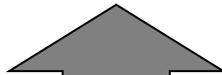

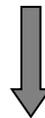
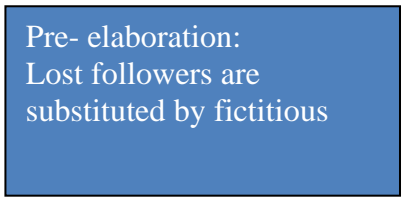
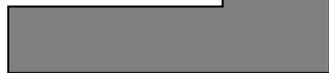
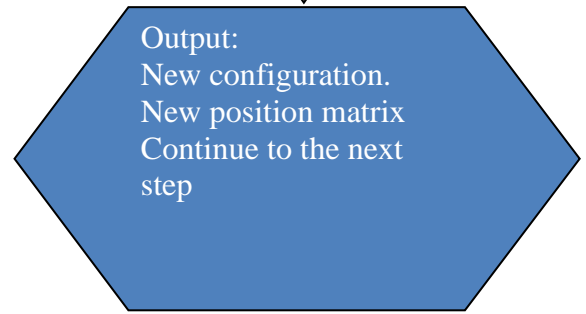

**Figure1** Flow chart of the particle position calculation

## 3. Applications and numerical results

In the previous papers [7],[14-15] we have investigated some applications of the algorithm. Now our intention is to test it in more complex cases also to highlight its limits. Limiting ourselves into two dimensional systems we try to change the lattice type, the neighbours and to use an ASTM shape sample instead of the usual simple form. We will impose a certain strain on the samples acting on the leaders. Our intention in these simulations is to show the importance of the lattice type and all the other features of the software tool on the obtained results. So far we shall consider shear tests, tensile tests and tensile tests for ASTM samples. For every test we shall show and discuss the movement of the particles, the XY movement of a particle of particular relevance (if present) and some pseudo energetic considerations.

Case a) Shear test
As first case we would like to consider a shear test and to evaluate the influence of the lattice type, of the interactions rule and of the second gradient neighbors on the deformation obtained. Therefore we consider a square specimen discretized by a square Bravais lattice. The specimen is subject to a shear with constant velocity 0.1 unit/time step in x axes, by the leaders. We consider 100 time step of strain. When we do not have specified interaction rules between the followers we use the barycenter rule i.e. the coordinate of the follower is computed as the gravity center of all its neighbors. Moreover, if not specified, the number of neighbors is given by the coordination number of the chosen lattice, while in second gradient there is a second shell. No fracture is considered in this case.
In Figure2 we can see the configuration of the lattice together with the PE2 contour plot, to compare contiguous configurations. The leaders are red, the followers blue and the frame is in orange colour. Lateral deformation are non linear and comparing contiguous configuration, by PE2 function, differences are larger close to the leaders. No deformation of the top and bottom line can be outlined

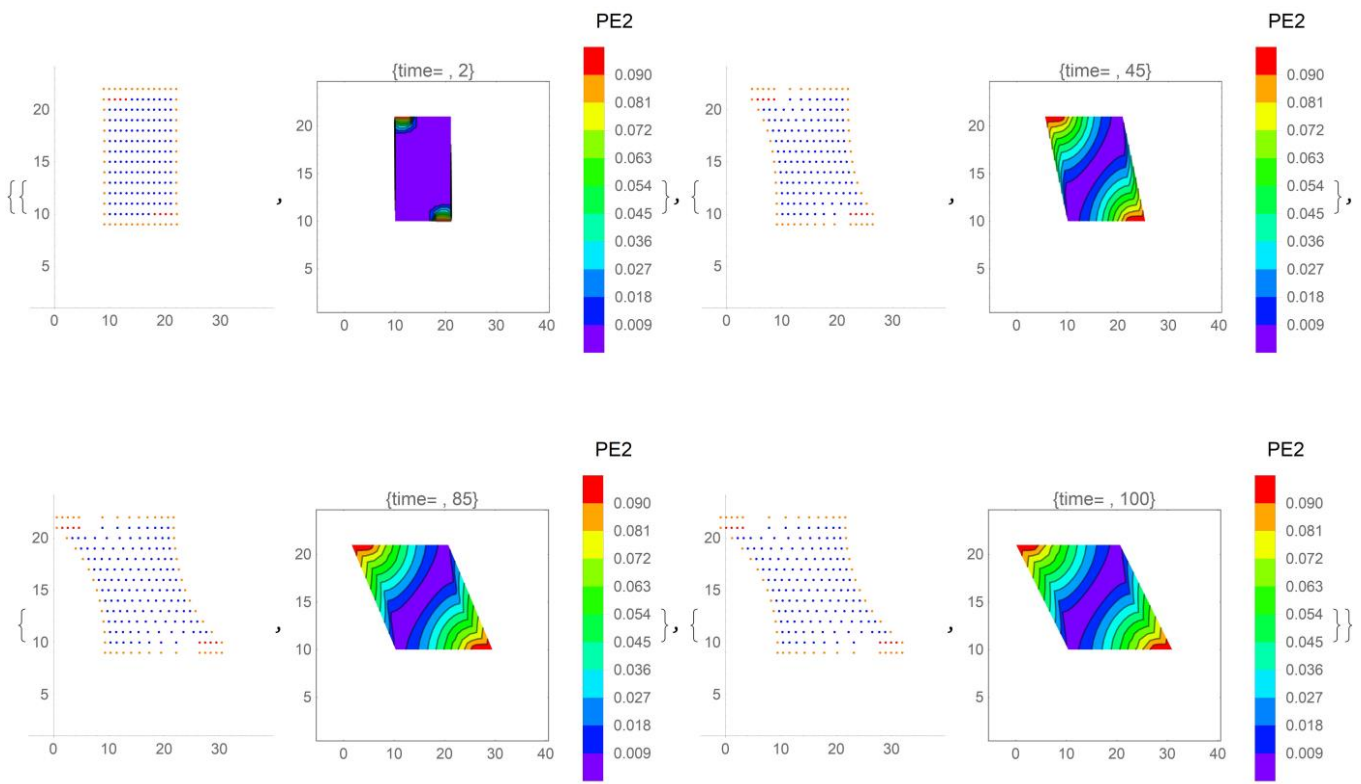

**Figure2** Configuration of the lattice over different time (2, 45, 85 and 100) in shear test square lattice together with PE2 contour plot, indicating differences between contiguous configurations.

A more marked lateral deformation curve can be seen if we use a honey comb lattice (see Figure3). This is an example of how different deformed configurations can be obtained by changing lattice type holding all the other conditions. We remark that this is not a Bravais lattice but we have used it owing to its large practical applications. As in the previous case PE2 contour plot shows that differences between contiguous configurations are larger close to the leaders

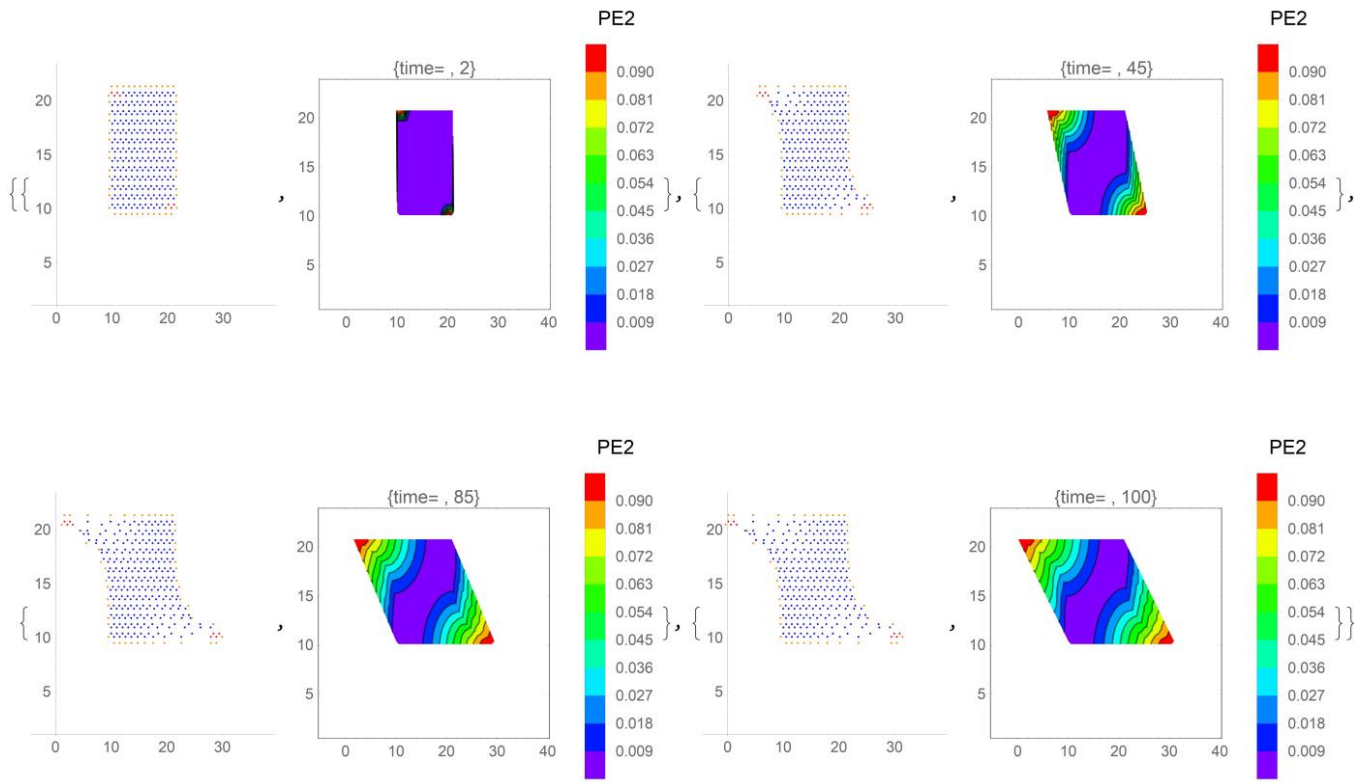

**Figure3** Configuration of the honey comb lattice over different time (2, 45, 85 and 100) in shear test square lattice together with PE2 contour plot, indicating differences between contiguous configurations.

A more interesting case, using a square lattice, is shown in Figure4. Here we have used a rule for the follower making use of "mixed coordinate", which means the y coordinate is dependent on the evolution of the x coordinate. This allows us to obtain lateral contraction, i.e. Poisson effect. The result of the shear test is a strange "window" flag. Once again PE2 contour plot show that large differences between time contiguous configurations can be outlined close to the leaders, from no particular differences with the preceding plots.

To stress the evolution of two symmetric points in y coordinate we consider point 4 and 133; we have chosen one above and one below the center line. The points are numbered as can be seen in Figure 5 and the evolution of the points over time is showed in Figure 6. Owing to the lateral contraction, the y coordinate of points above this axis decrease, with shear, while below they are increased. As can be seen it is not a simple lateral contractor, as in the previous work, but a more complex behavior owing to the lattice. Note the delay reaction time, because the involved points have to be informed about the displacement of the leaders.

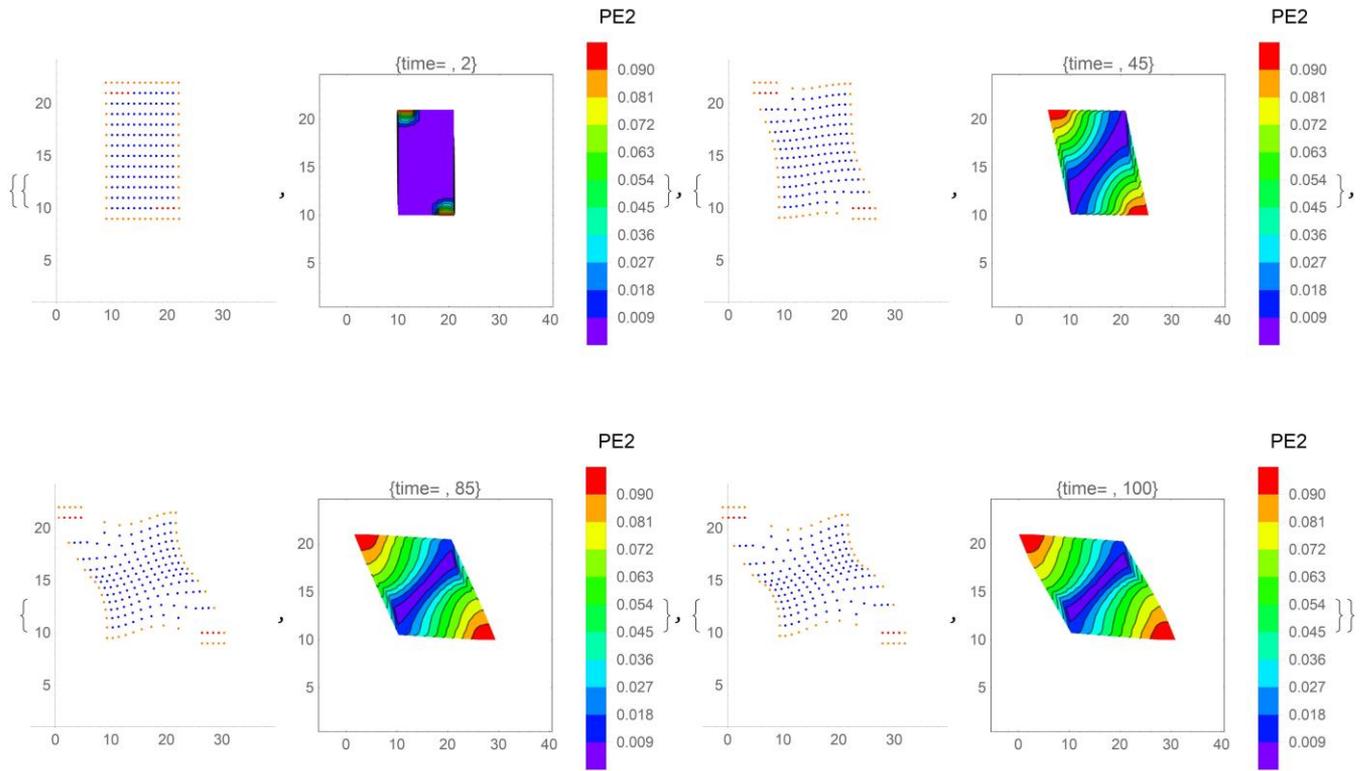

**Figure4** Configuration of the lattice with Poisson effect over different time (2, 45, 85 and 100) in shear test square lattice together with PE2 contour plot, indicating differences between contiguous configurations.

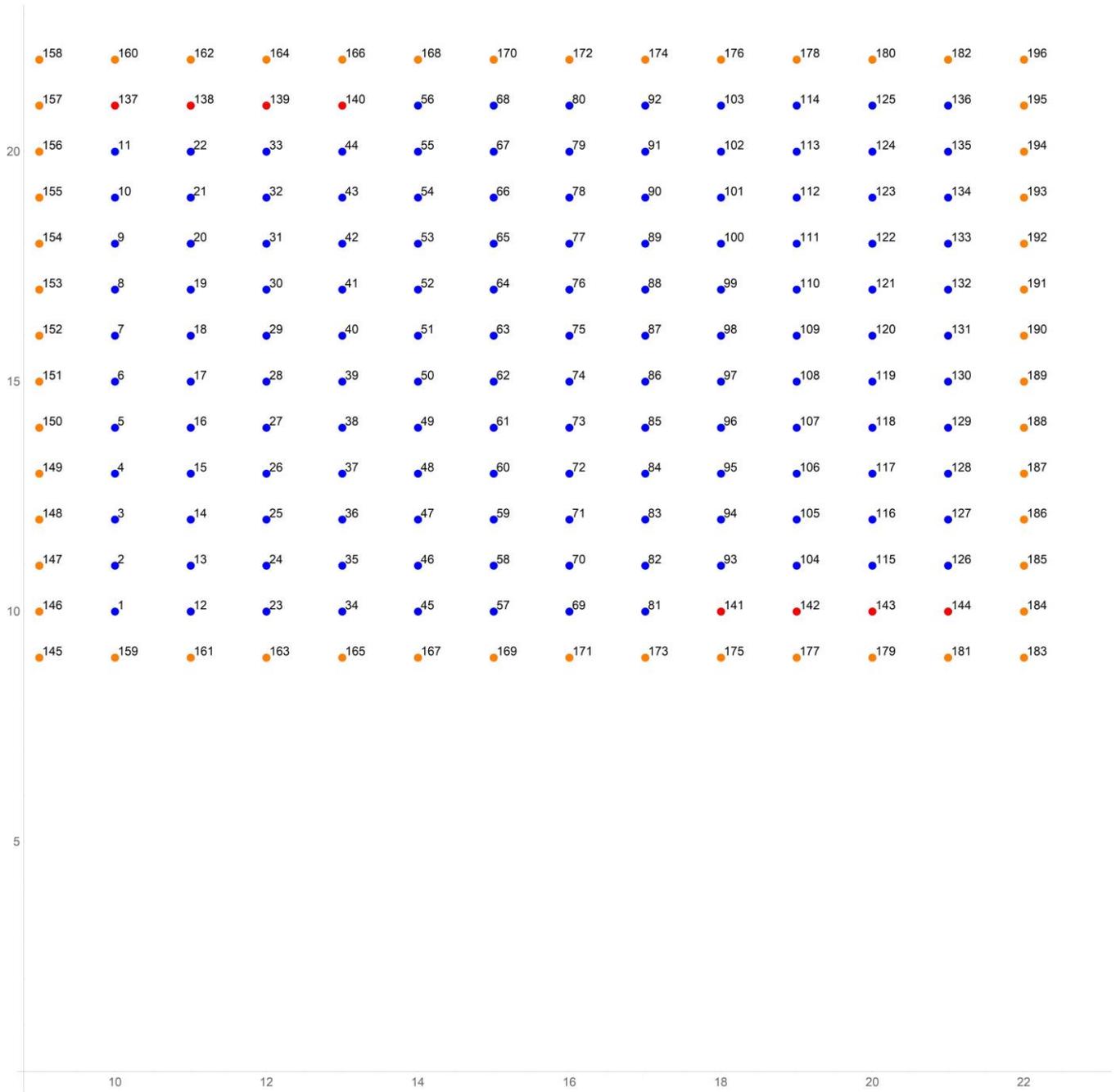

**Figure 5** Numbered lattice. Red points are the leaders, yellow the frame and blue the followers.

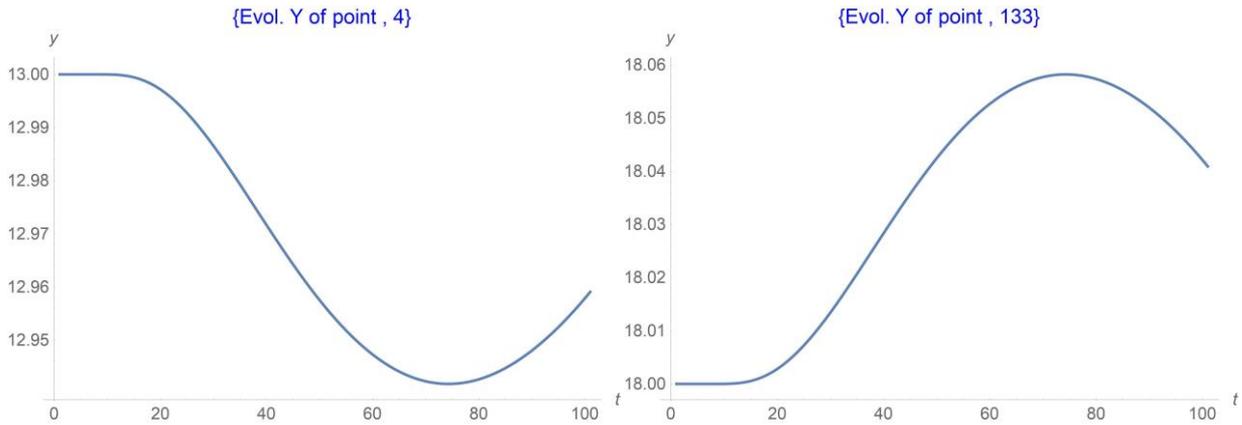

**Figure 6** Evolution of Y coordinate of the point j=4 and j=133 versus time

Quite similar behavior can be observed if we use second gradient model, changing the shell of neighbors (see Figure7 and Figure8 ). Differences are in a more stiff reaction, owing to the larger numbers of neighbors involved in calculating the follower's positions. This can also be seen in Figure8 where a smaller excursion of the Y coordinate can be evaluated than before.

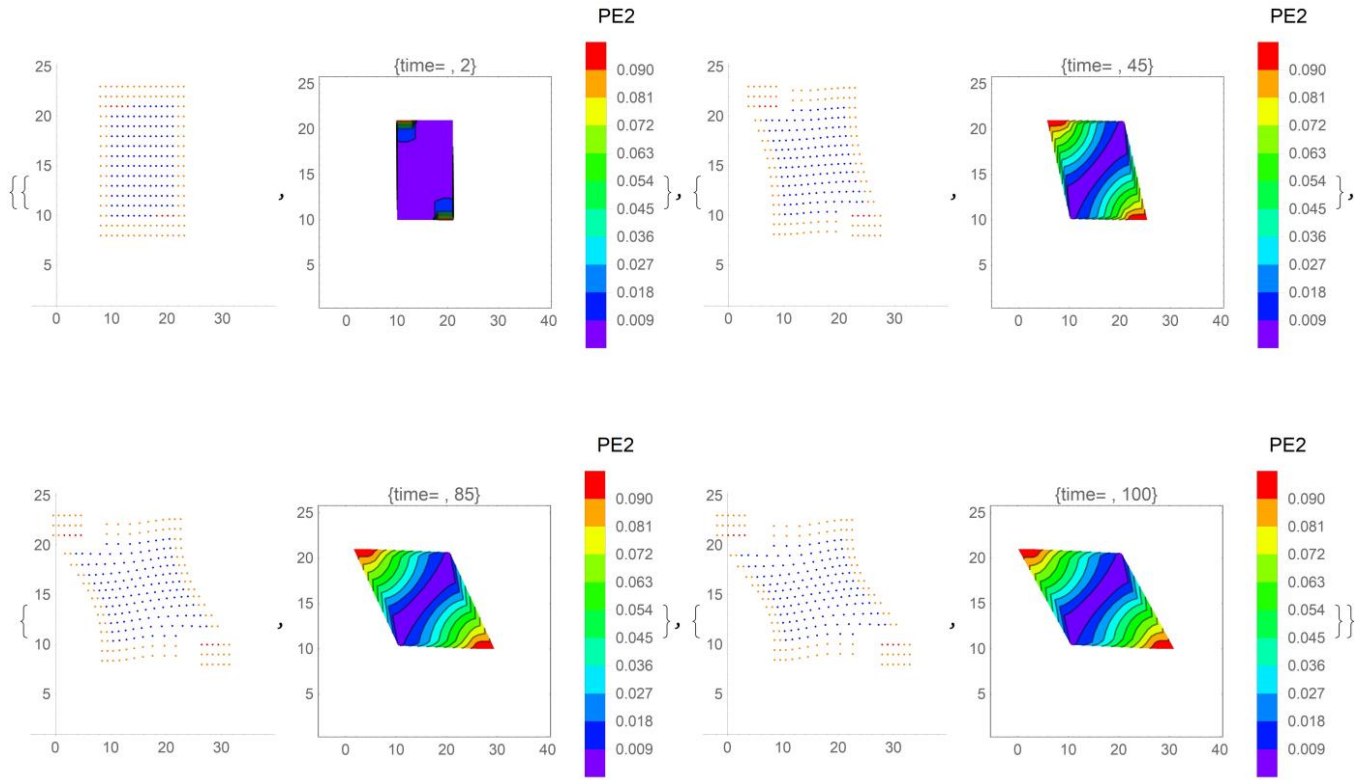

**Figure7** Configuration of the lattice with Poisson effect over different time (2, 45, 85 and 100) in shear test square lattice, second gradient, together with PE2 contour plot, indicating differences between contiguous configurations.

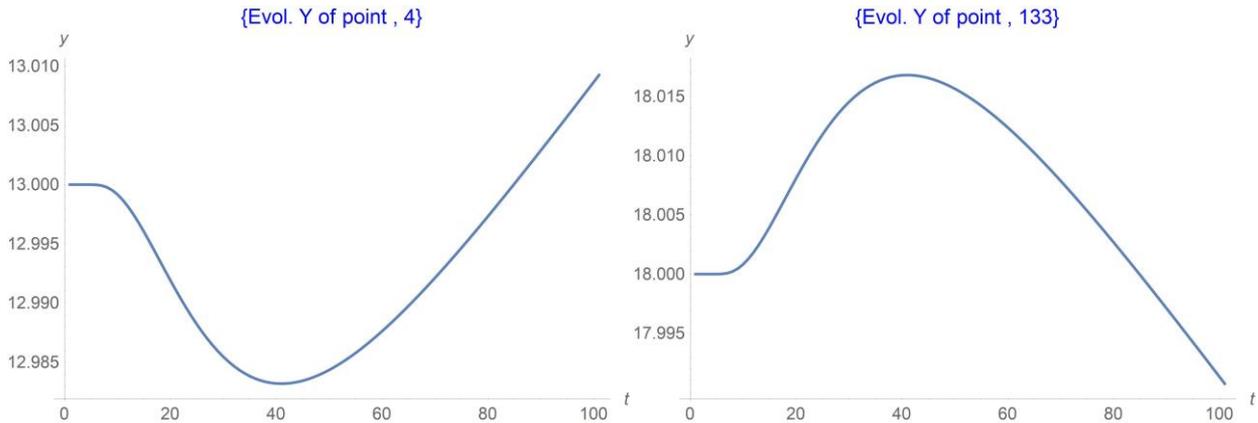
**Figure8** Evolution of Y coordinate of the point j=4 and j=133 versus time (Second gradient case)

Finally in Figure9 fracture mechanism of simple square lattice is shown. We go back to the first simulation, with a square lattice but we have chosen a fracture distance of 2.5 units and the fictitious have been posed in neutral position, as explained in the previous works [7,14,15]. It can be noted that the leaders bring with them some of the followers; this depends on a complex balance between the leader's attraction and the resistance offered by the followers. Changing condition results in changing the number of the "attached" followers. After the fracture the particles return back to their equilibrium position. Note that if we would position the fictitious in another location we would obtain a different result. Pseudoenergy has symmetric behavior, as expected. Remember that the pseudoenergy concept was considered on not fractured sample, and it is not calculated on the fictitious points but on the followers so it is not significant.

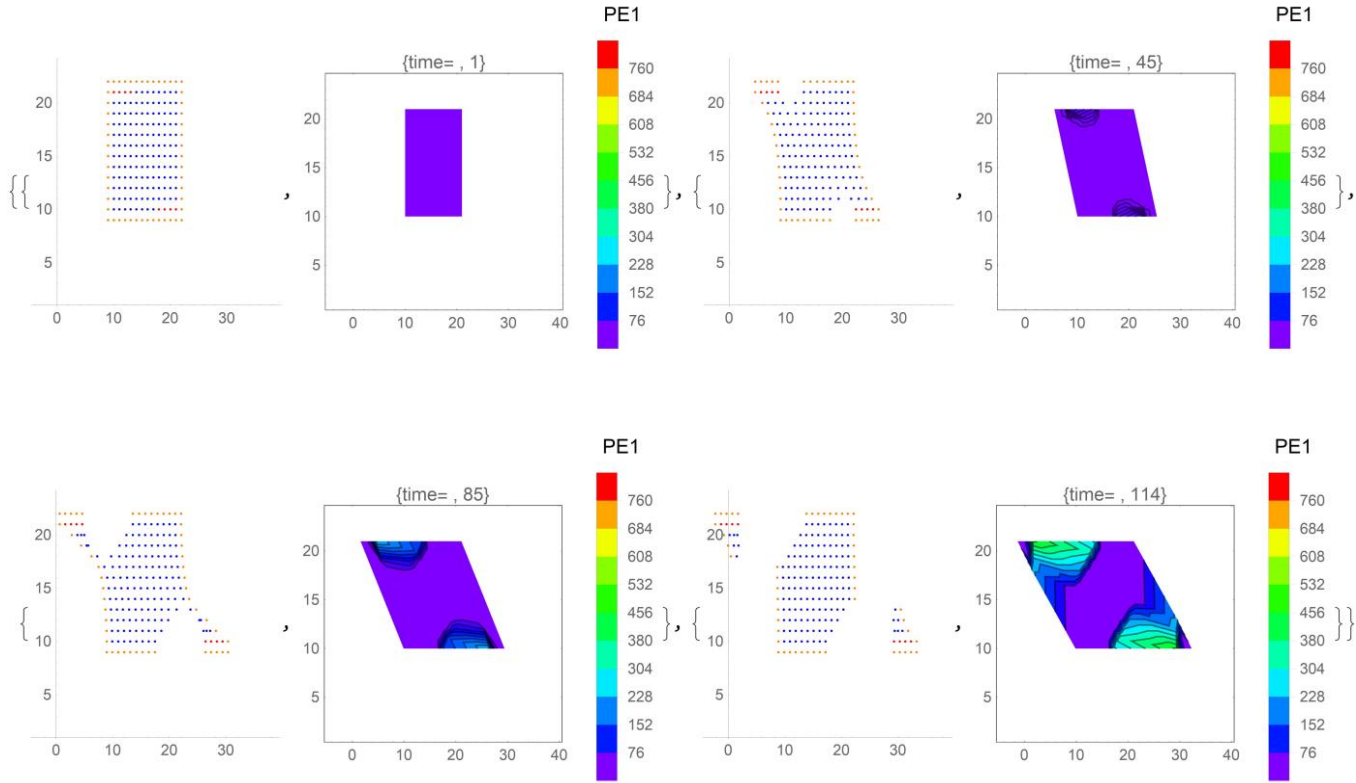

**Figure9** Shear Configuration of the lattice over different time (2, 45, 85 and 114) in shear test square lattice with fracture together with PE1 contour plot, of pseudoenergy.

Case b) Tensile test

For the next example we shall consider a square sample undergoing different cases of tensile tests; the aim of these tests is to stress the importance of the chosen lattice and of the rules, determining the follower's motion, on the deformation. In Figure10 we are showing configuration and PE1 pseudoenergy values of the first test. Fracture distances are 6 units and speed is 0.5 units/step, for 150 steps long. When a distance between the points is larger than fracture distances the sample is broken and the followers go to equilibrium position; if no followers remain attached to the leaders (it depends on the distances, we shall see later in other cases) they return to their initial position. As explained in a preceding work [7] the convexity, in the fracture mechanism, is related to the presence of the frame. PE1 plot show as, before fracture, there are areas of stress concentration. Higher stress areas are close to the leaders. The trend of this point (see Figure11, where the X evolution on time of a central point is shown) is quite linear during traction but it becomes non linear when the followers remain alone and return back. This because the traction is imposed with constant speed, while the reassembly of the points is driven by the follower's rule. Note the "hesitation area" close to the peak of the curve.

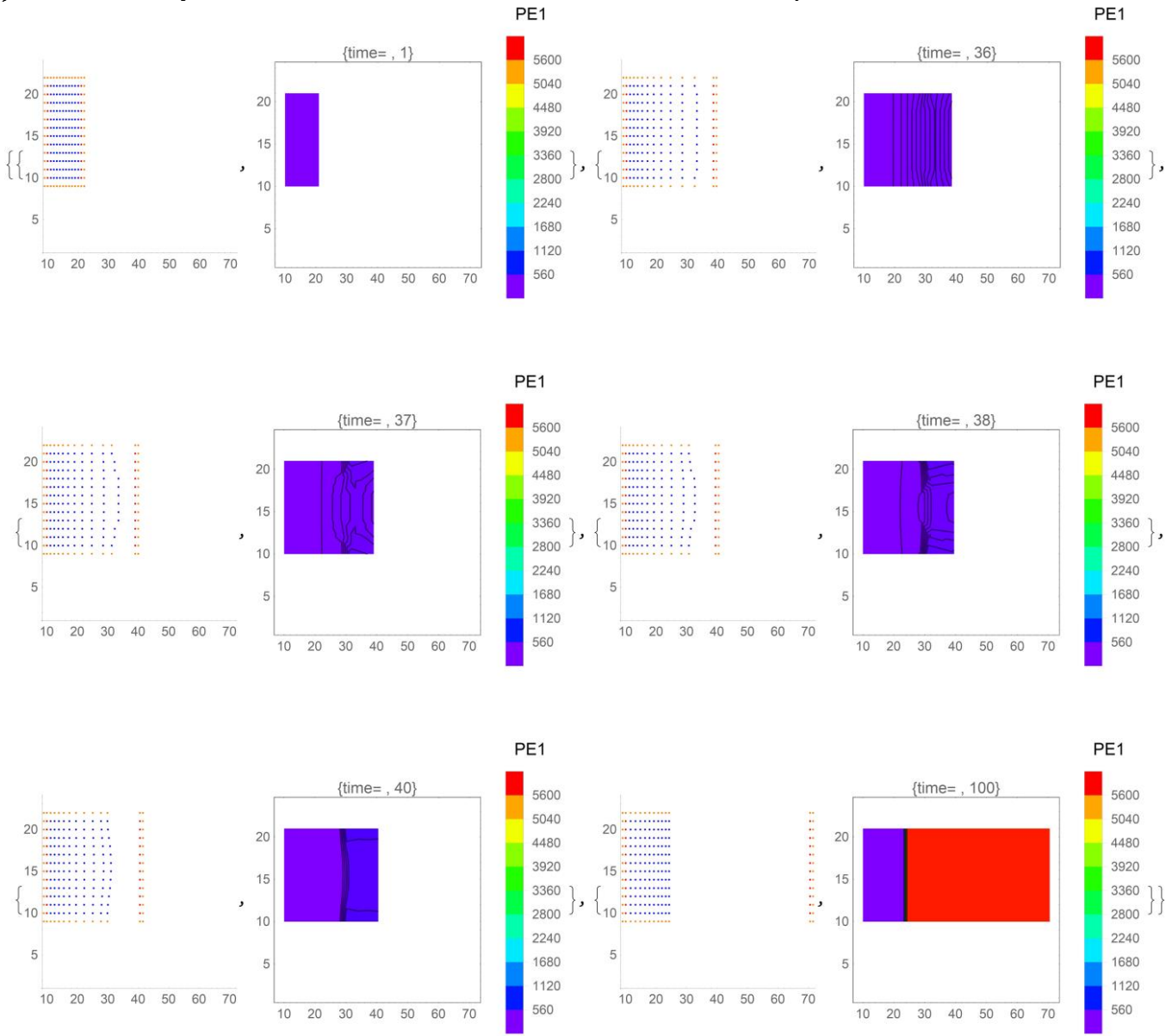

**Figure10** Tensile test with fracture square lattice. Configuration over different time (1, 36, 37, 38, 40 and 100) together with PE1 contour plot, of pseudoenergy.

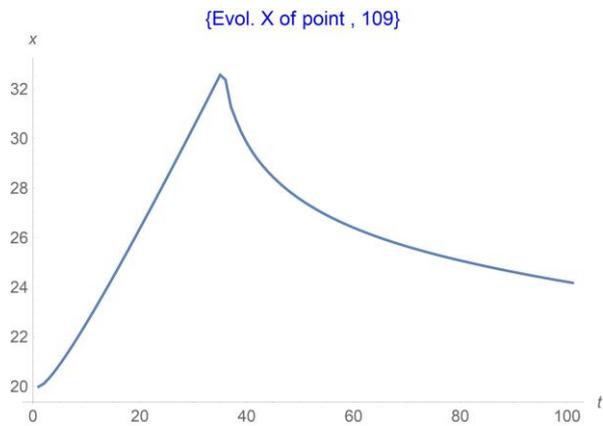

**Figure11** Evolution X coordinate of the point j=109 versus time

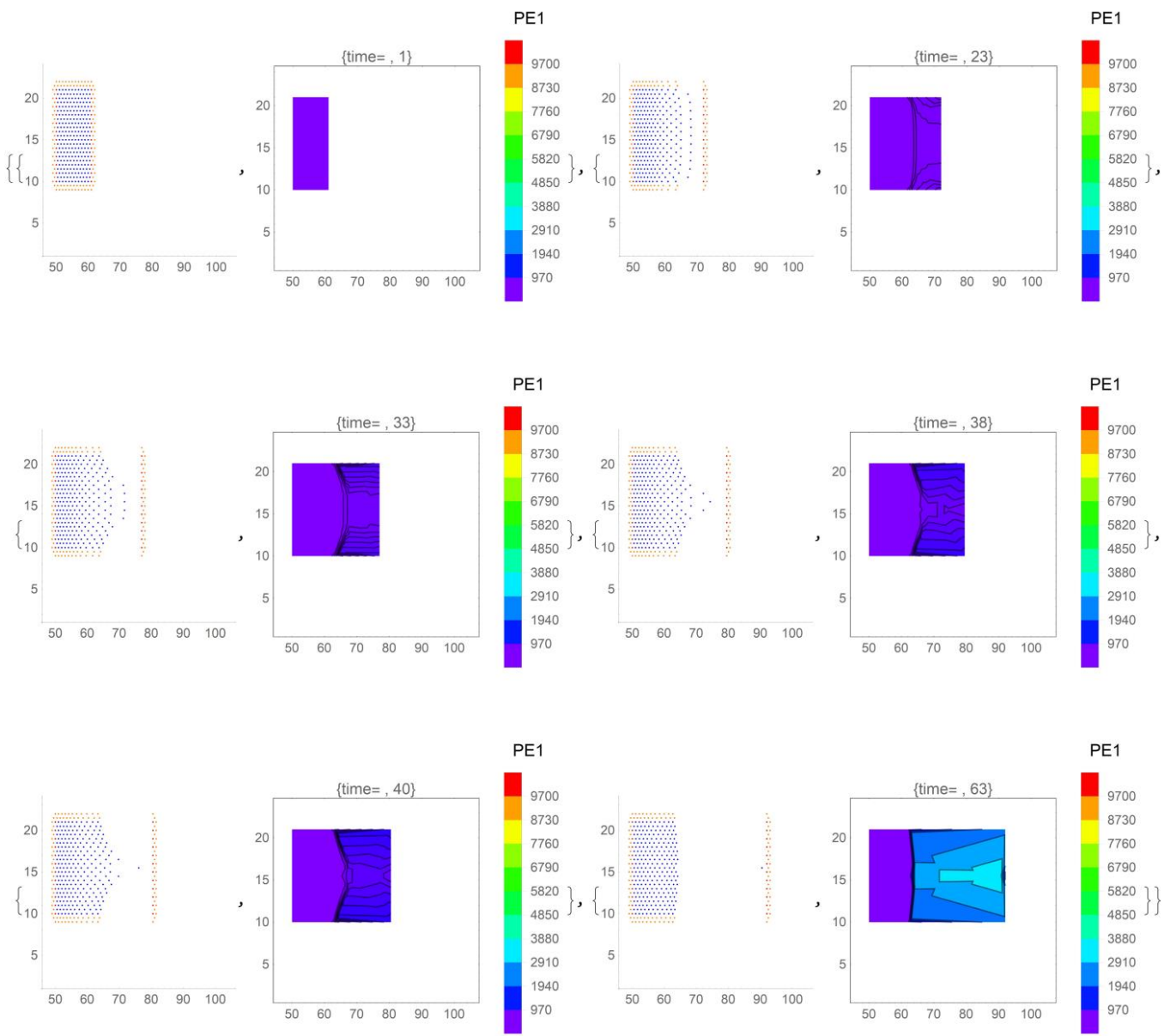

**Figure12** Tensile test with fracture rectangular lattice in second gradient. Configuration over different time (1, 23, 33, 38, 40 and 63) with PE1 contour plot, of pseudoenergy.

The importance of the lattice can be seen in Figure12 and Figure13 where the same test is computed but using a rectangular centred lattice and second gradient interaction. The fracture mechanism is quite different together with the final configuration. Only the central point remains attached to the leaders, because it suffers the strongest attraction. This example shows, once again, that change in model parameters lead to different behaviours.

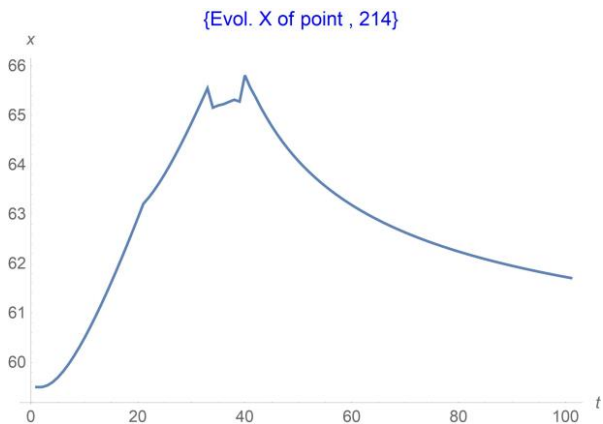

**Figure13** Evolution X coordinate of the point j=214 versus time; after the fracture they return to the original position but there is a transient period before relaxation.

In Figure13 the evolution of the x- coordinate for a central point close to the leader's line is shown. It can be noted that, after the fracture, there is a complex movement before the relaxation curve.

Another example can be obtained if we consider the same conditions as before but we change the neighbour's number to five and consider a first gradient interaction; we obtain a completely different result. In Figure14 we have the same test as before; the lower number of particles involved in the calculation of the relative position makes the sample much more fluid, allowing detachment of a larger number of particles, as we can see on the right side of the pictures. The fracture mechanism also is different with respect to the preceding case.

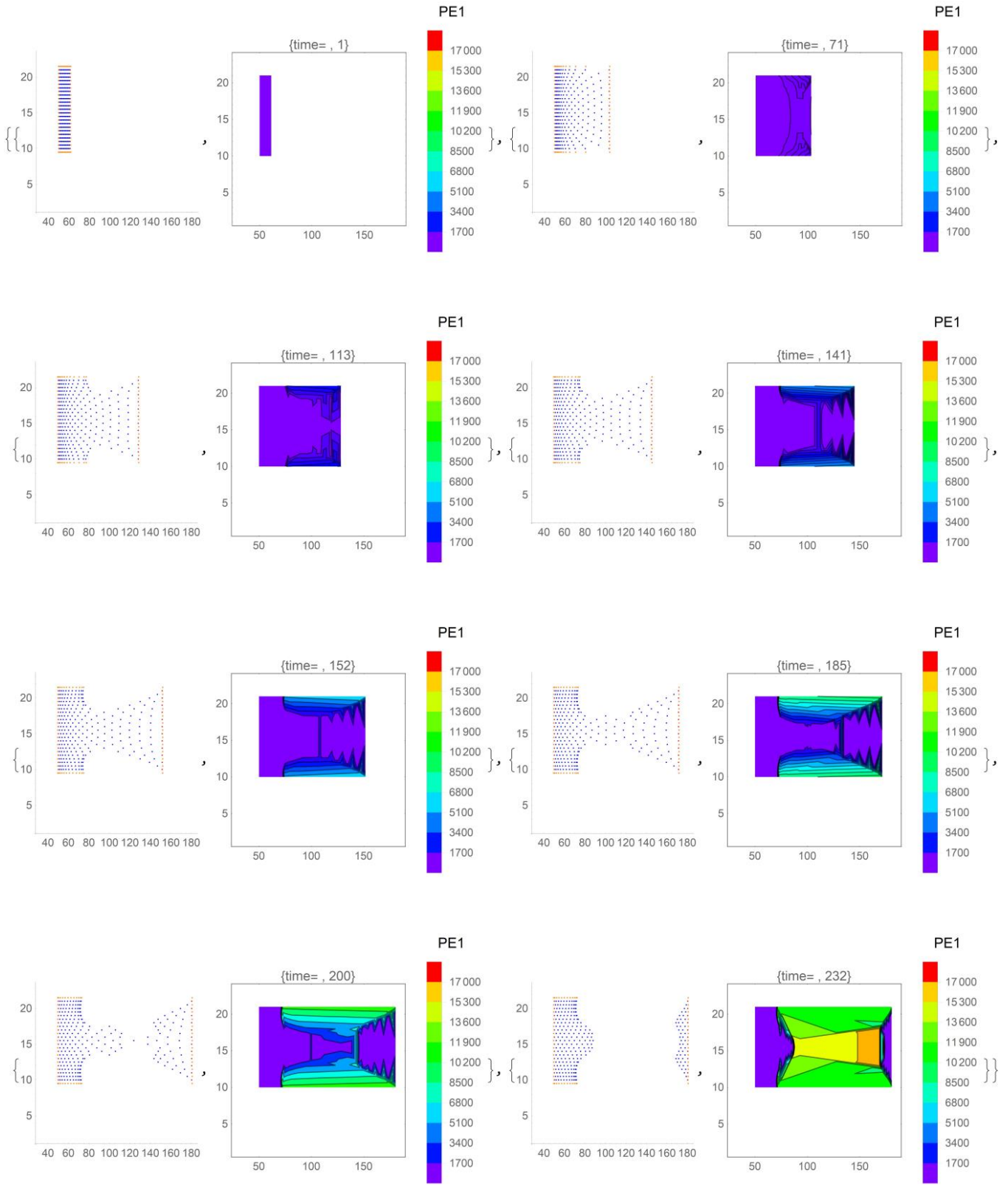

**Figure14** Tensile test with fracture rectangular centred lattice, coordination number 5. Configuration over different time (1, 23, 33, 38, 40 and 63) with PE1 contour plot, of pseudoenergy.

Another example of a different fracture mechanism can be outlined in Figure15 where a hexagonal lattice, always in the same condition, has been used. Once again we obtain a different number of detached followers that remain close to the leaders, and a different final configuration. The PE1 plot shows a simple behaviour.

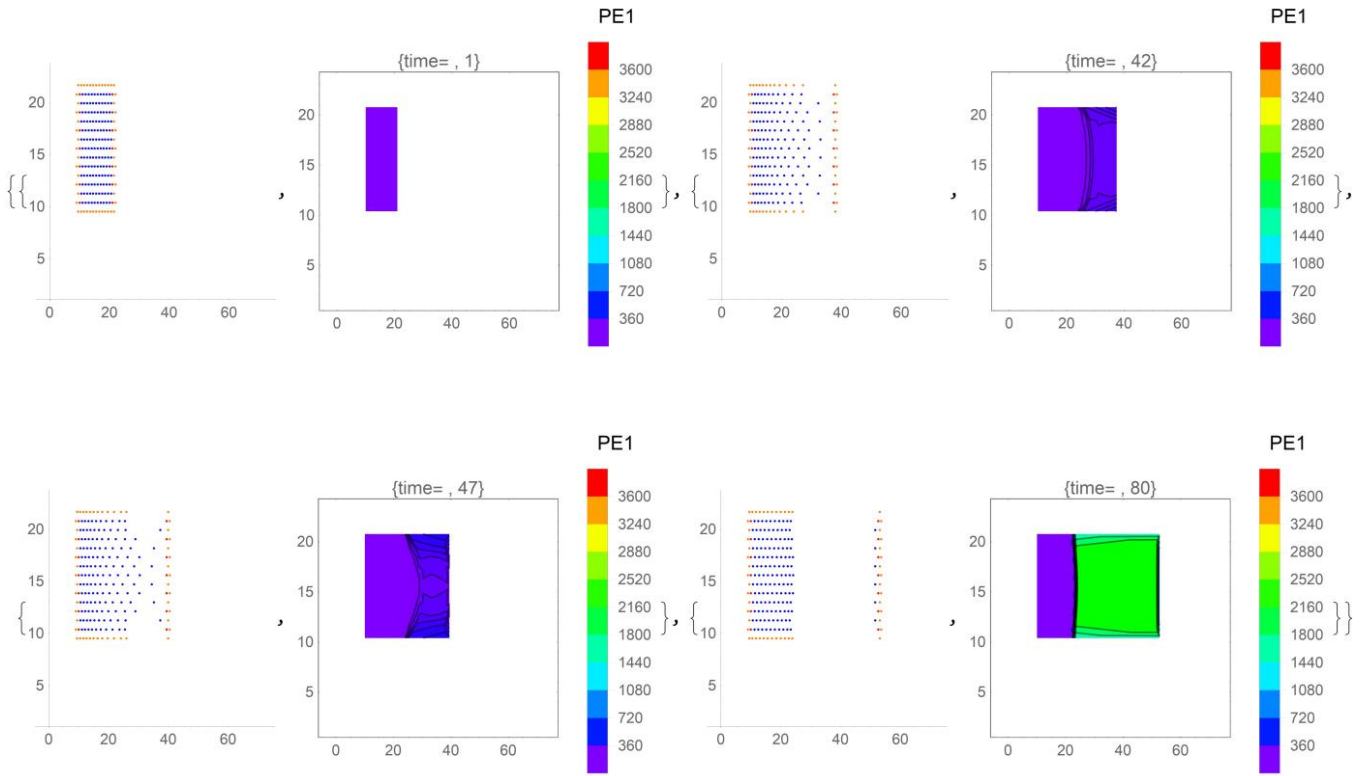

**Figure15** Tensile test with fracture hexagonal lattice. Configuration over different time (1, 42, 47 and 80) with PE1 contour plot, of pseudoenergy.

An interesting phenomenon can be seen if we consider an oblique lattice Owing to the asymmetry (see Figure16; look at the five red leaders on the right) of the leaders with respect to the frame a particular breakage fracture can be observed (see Figure17). In fact if we consider a symmetry axes in x direction we can note two leaders close to the frame in the upper level and only one close to the bottom. This leads to a fracture starting from the bottom where the attraction of the leaders is lower. It seems to rip a piece of paper. The fracture distance is 10 units and the speed is 0.4 unit/time step.

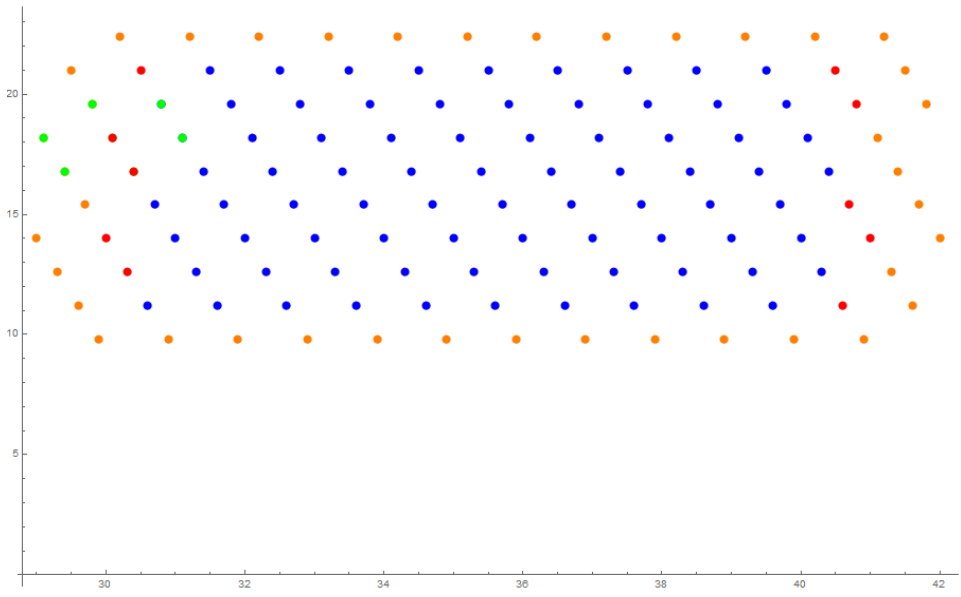

**Figure16** Oblique lattice tensile test. Red points are leaders, blue point followers and yellow the frame. First gradient case. Note the asymmetry of the leaders with respect of the frame

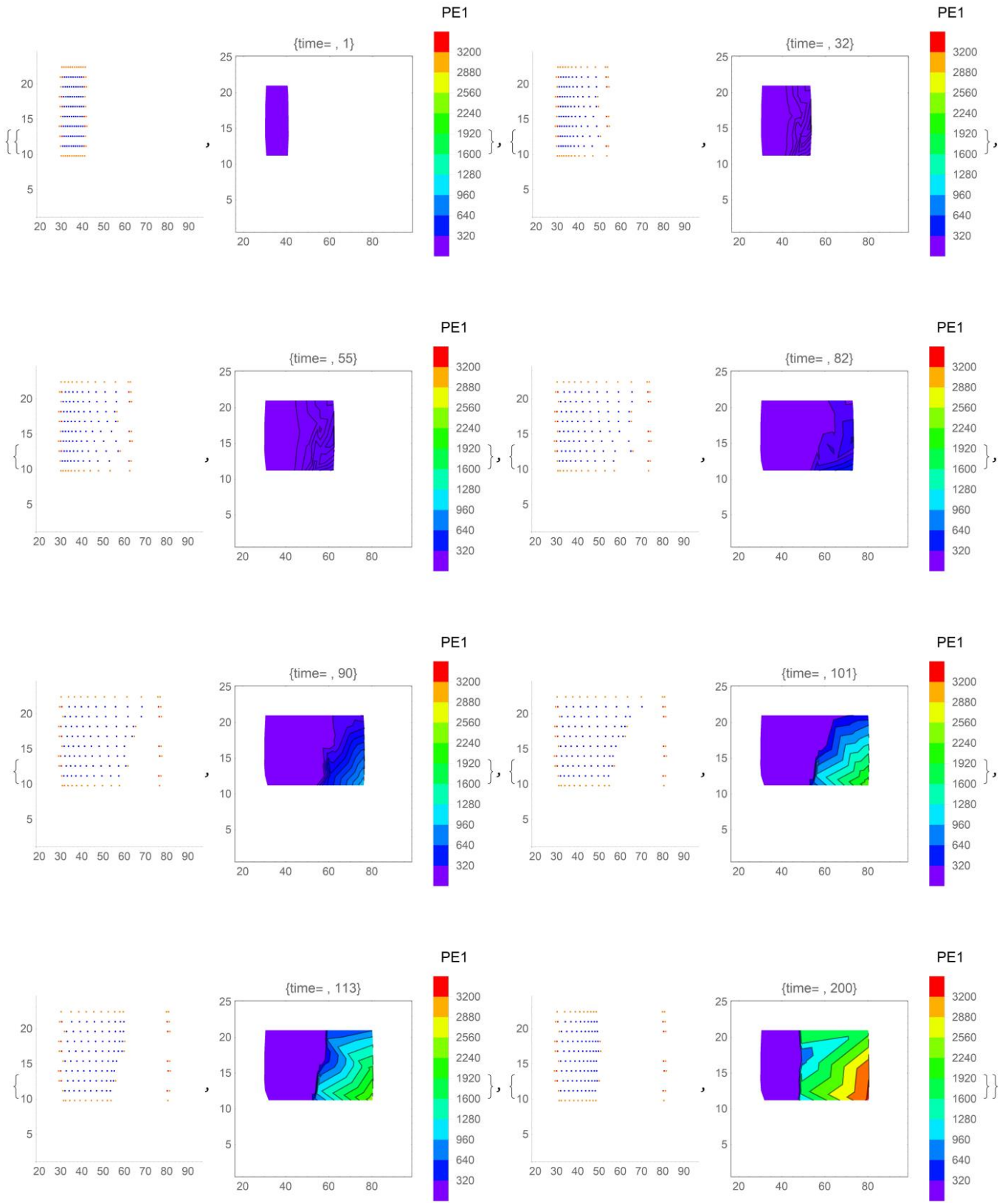

**Figure17** Tensile test with oblique lattice: breakage fracture. Configuration over different time (1, 32, 55, 82, 90,101,113 and 200) with PE1 contour plot, of pseudoenergy

The last lattice we have investigated has been the honey comb. In Figure18 and Figure19 a fracture tensile test is shown in first and second gradient case. The first one shows a very brittle fracture, quite similar to the square lattice but without the small convexity of the fracture line. On the contrary the second gradient shows a very pronounced convexity. This so different behaviour suggests to us, once again, the needs to investigate better the rules that determine the follower positions, are linked to the observed results: this will be the object of a subsequent paper.

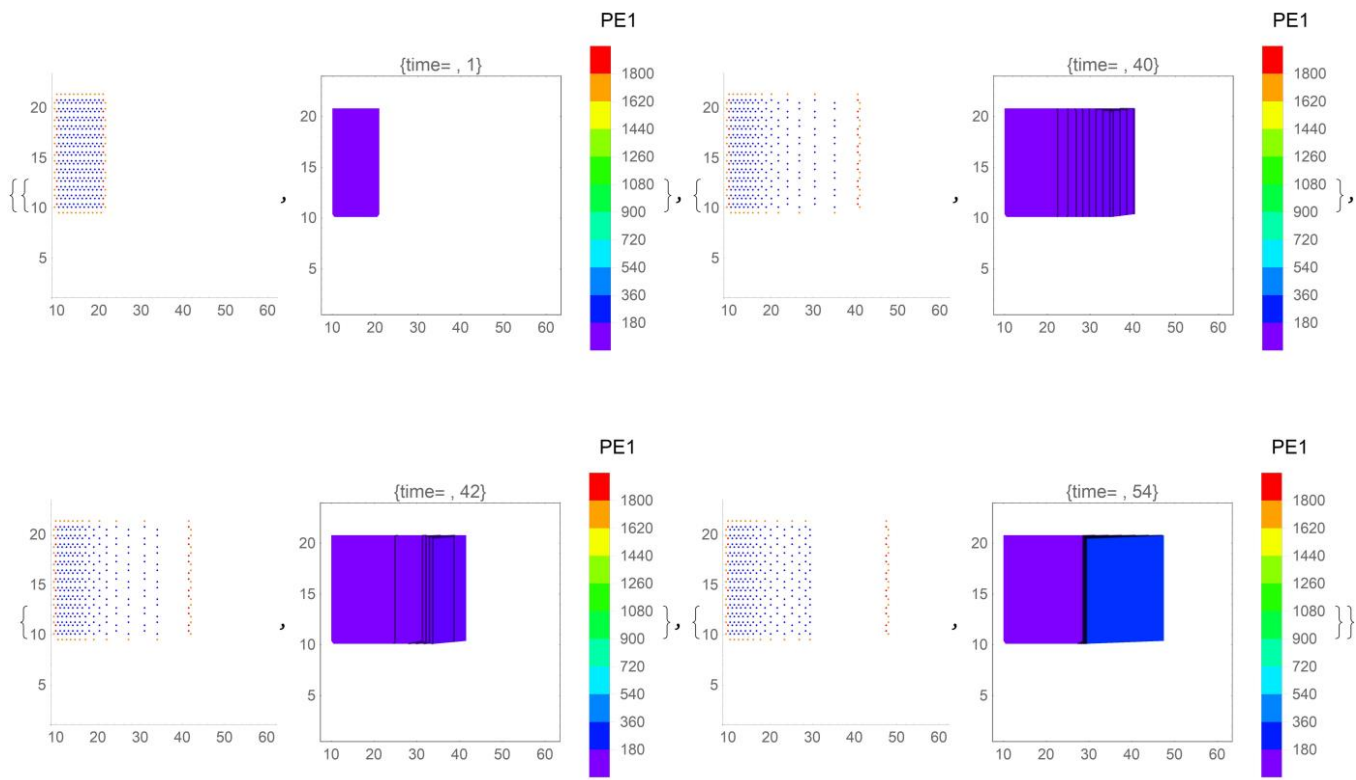

**Figure18** Fracture tensile test honey comb lattice. First gradient case. Configuration over different time (1, 40, 22, and 54) with PE1 contour plot, of pseudoenergy

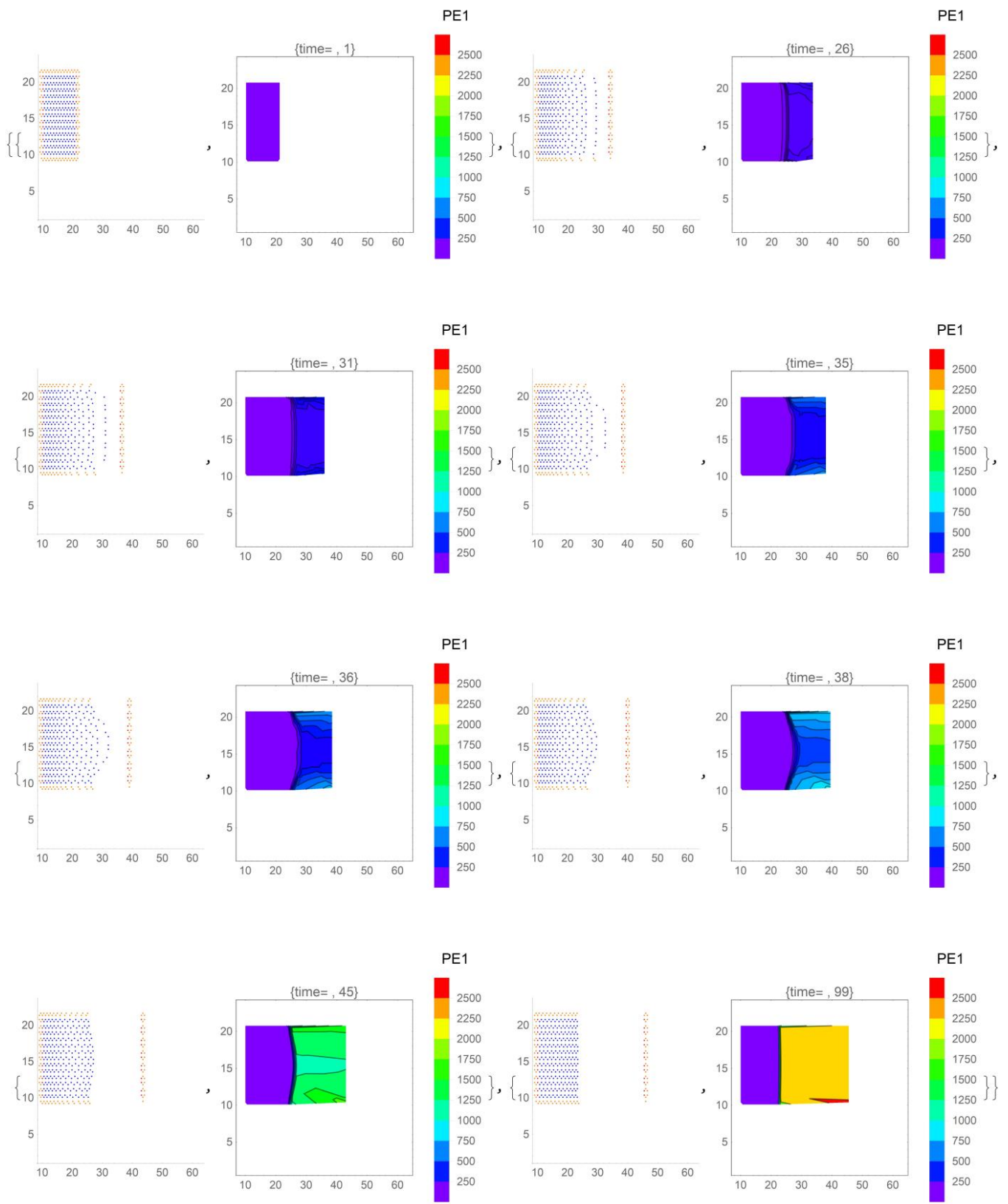

**Figure19** Fracture tensile test honey comb lattice. Second gradient case. Configuration over different time (1, 40, 22, and 54) with PE1 contour plot, of pseudoenergy

Case c) ASTM test

In order to investigate sample behavior in a well known shape we have carried out some calculations considering a specimen quite similar to the ASTM D638 standards for tensile tests. The specimen is clamped at both ends on a surface and pulled on one side, so, in this case, the leaders are many; the test speed is of 2.5 unit/step, in x positive direction, for 150 traction step e 2500 relaxation step. We shall consider three different cases.

1. Simple tensile test, where we compute deformation for square lattice, a rectangular centered lattice, rectangular centered lattice with neighbor's number reduced to five and honey comb lattice.
2. Poisson case, where we modified interaction rule to obtain lateral contraction during elongation; we compute deformation for square and rectangular centered lattices with reduced neighbor's number to five.
3. Finally the fracture case is investigated for the rectangular centered lattice with coordination number five.

So let's start by considering the simple tensile test (see Figure20). A larger number of time steps is required for relaxation, owing to the larger number of points used to describe the specimen; this does not mean a longer relaxation time, because unit time is arbitrary, only because the influence of the displacement propagates at one shell (first gradient case) each time step we need many steps to involve the whole sample. Pseudoenergetic plot show as higher level of the parameter is reach during elongation when point's distances are larger as expected.

It should be noted that the sample does not reach a symmetric final configuration as we can expect (see last figure); we have also wait for 10000 time relaxation steps without modification. On the contrary if we use a rectangular shape sample it does (the points are equally spaced) as can be seen in Figure21. There is no physical reason for this, our opinion is that this effect is linked to the particular equilibrium condition generate by the geometry. Figure22 shows that final configuration is more similar to a symmetric one, in second gradient case, owing to the larger number of points involved in the computation. We are working on this and on higher gradient computations.

If we consider the four simple tensile tests (Figure20, Figure23, Figure24, and Figure25 respectively) little quantitative differences in point distribution can be observed during the classical elongation of the specimen in the four cases. We can observe differences in the internal distribution on the points and in the convexity of the propagation front of the deformation i.e. see the convexity of the points between Figure20 and Figure23.

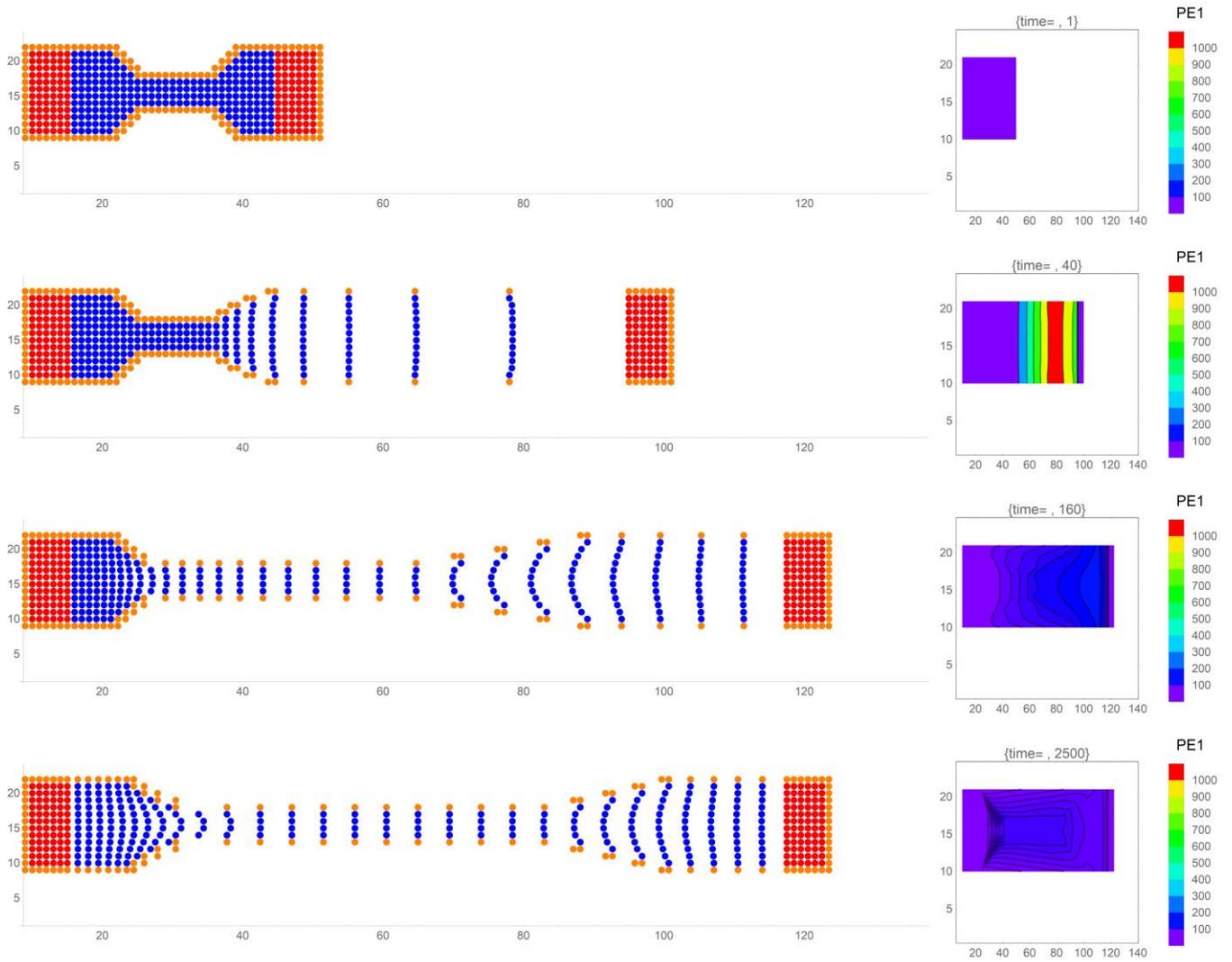

**Figure20** ASTM tensile test square lattice. Configuration over different time (1, 40,160, and 2500) with PE1 contour plot, of pseudoenergy

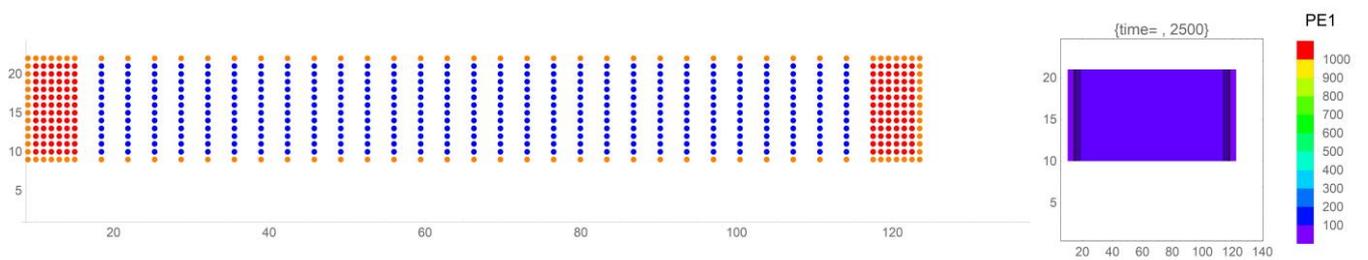

**Figure21** ASTM tensile test square lattice rectangular shape. Configuration at time 2500 with PE1 contour plot, of pseudoenergy

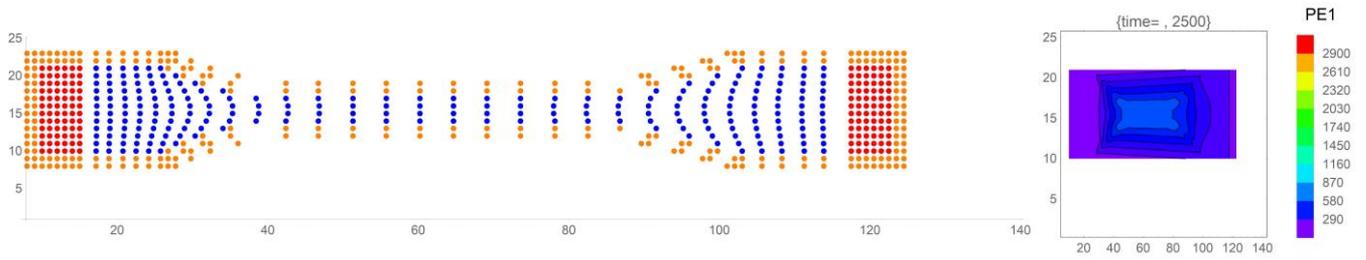

**Figure22** ASTM tensile test square lattice (second gradient). . Configuration at time 2500 with PE1 contour plot, of pseudoenergy

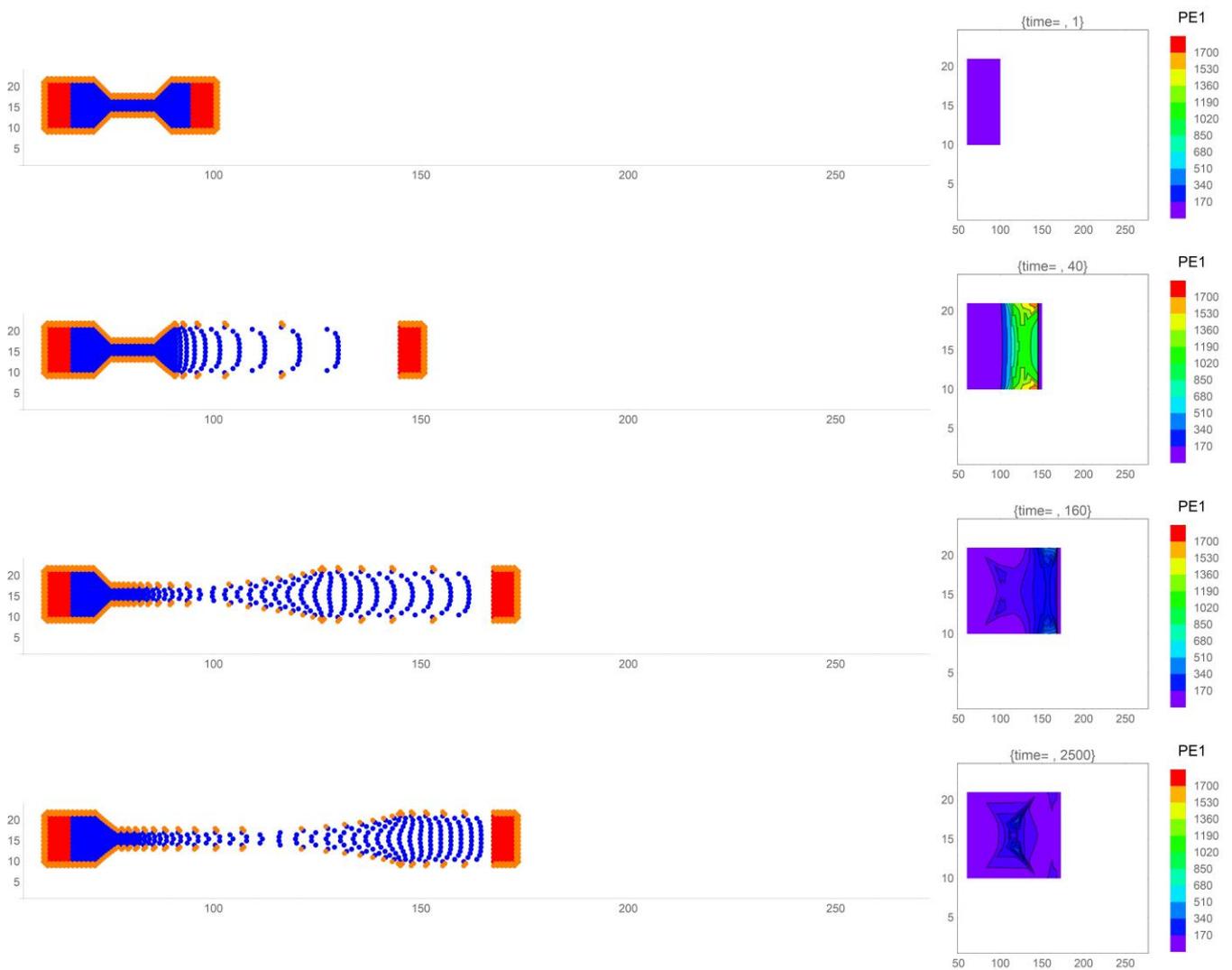

**Figure23** ASTM tensile test rectangular centred lattice. Configuration over different time (1, 40,160, and 2500) with PE1 contour plot, of pseudoenergy

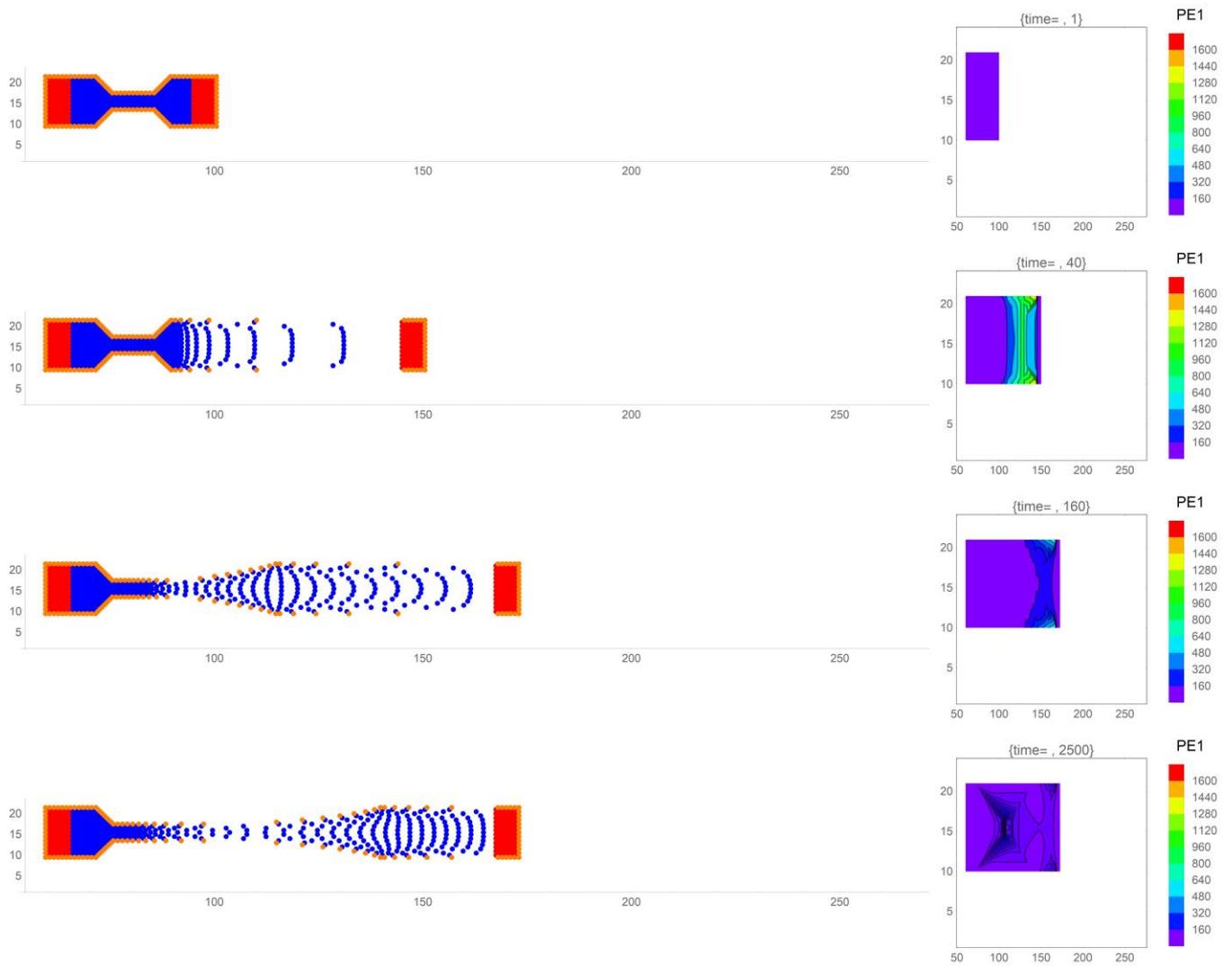

**Figure24** ASTM tensile test rectangular centred lattice (coordination number 5). Configuration over different time (1, 40,160, and 2500) with PE1 contour plot, of pseudoenergy.

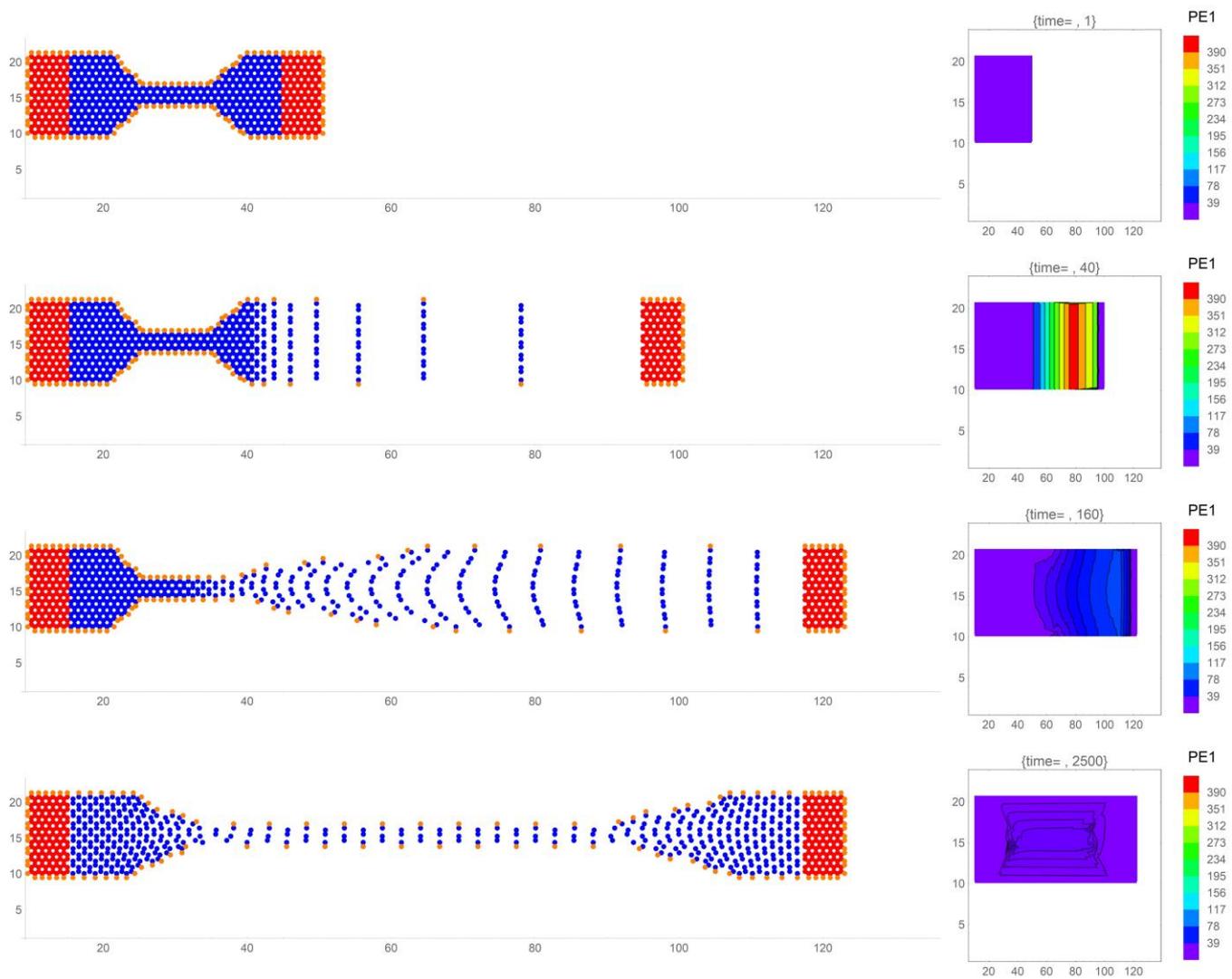

**Figure25** ASTM tensile test rectangular honey comb lattice. Configuration over different time (1, 40,160, and 2500) with PE1 contour plot, of pseudoenergy.

In the two cases we are considering the Poisson effect (see Figure26) it is possible to see lateral contraction. It seems the points cluster to create islands but this effect must be investigated better. In case of second gradient interaction this does not occur, as can be seen in Figure27.

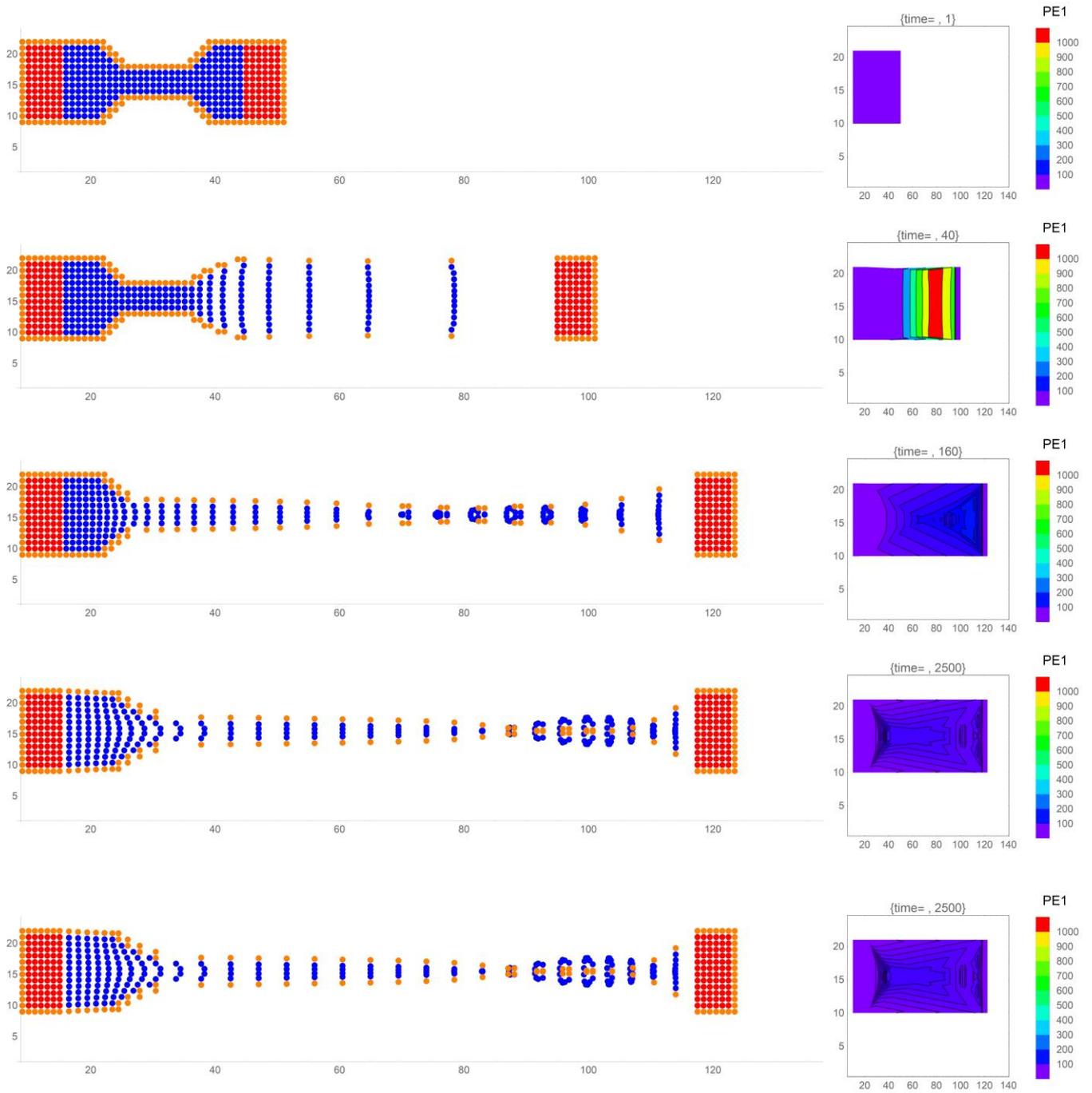

**Figure26** ASTM tensile test square lattice with Poisson effect. Configuration over different time (1, 40,160, and 2500) with PE1 contour plot, of pseudoenergy.

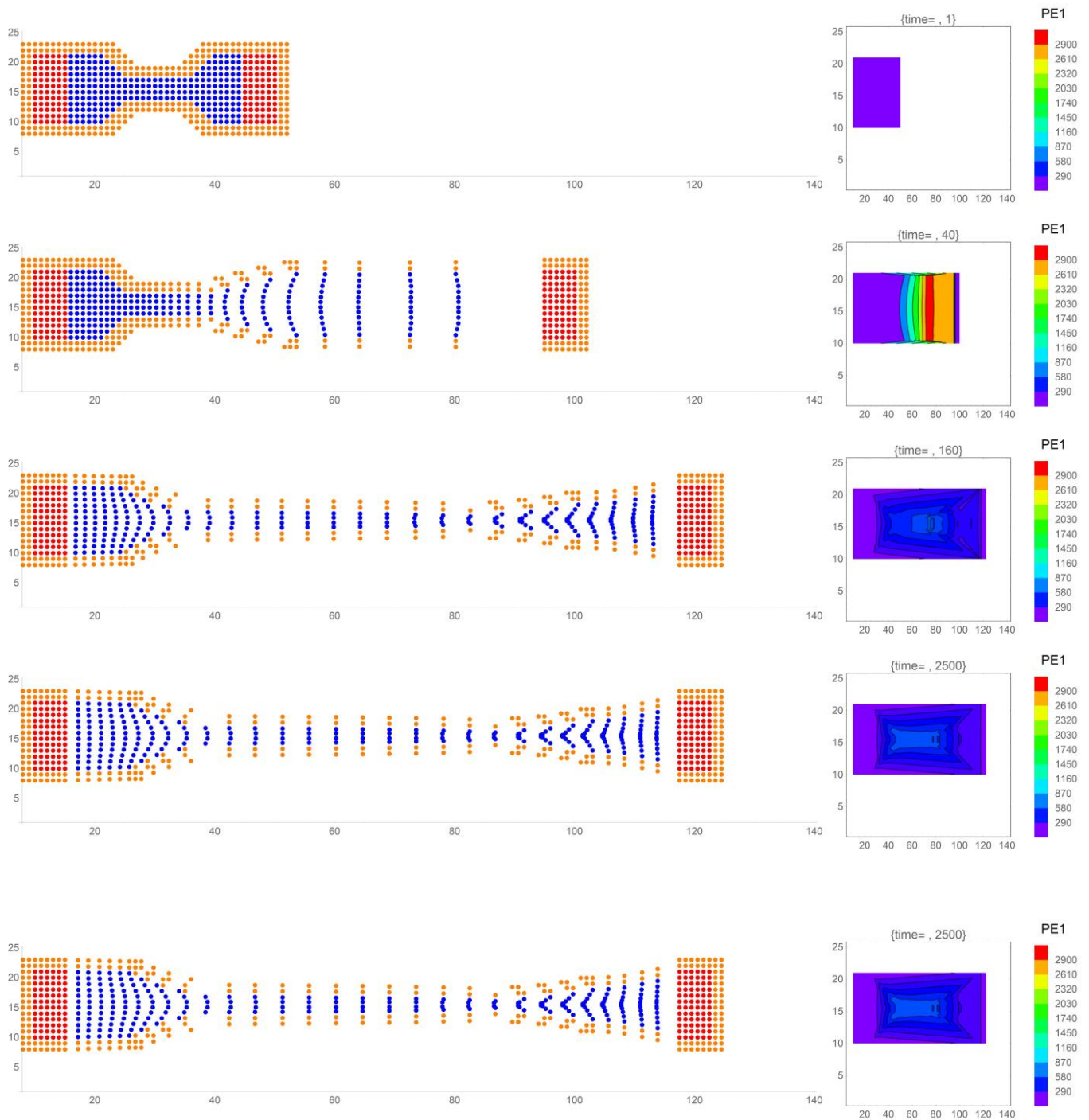

**Figure27** ASTM tensile test square lattice with Poisson effect (second gradient). Configuration over different time (1, 40,160, and 2500) with PE1 contour plot, of pseudoenergy.

To understand better the point's collapse phenomenon we consider a rectangular centered lattice with a reduced number of neighbors to five; we decide to reduce the neighbor's number to give greater ease of movement to the single point to enhance the phenomenon. In fact the collapse of the points occurs (see Figure28) crossing the central line and making cluster formation quite difficult to explain; it seems similar to the yielding phenomenon (whitening) observed in tensile test of polypropylene [8]. This behavior sounds quite unphysical and we remark the need to be connected with the constitutive equations of the materials we are investigating. We can obtain also auxetic behavior (negative Poisson coefficient) but not interpenetration of the particles.

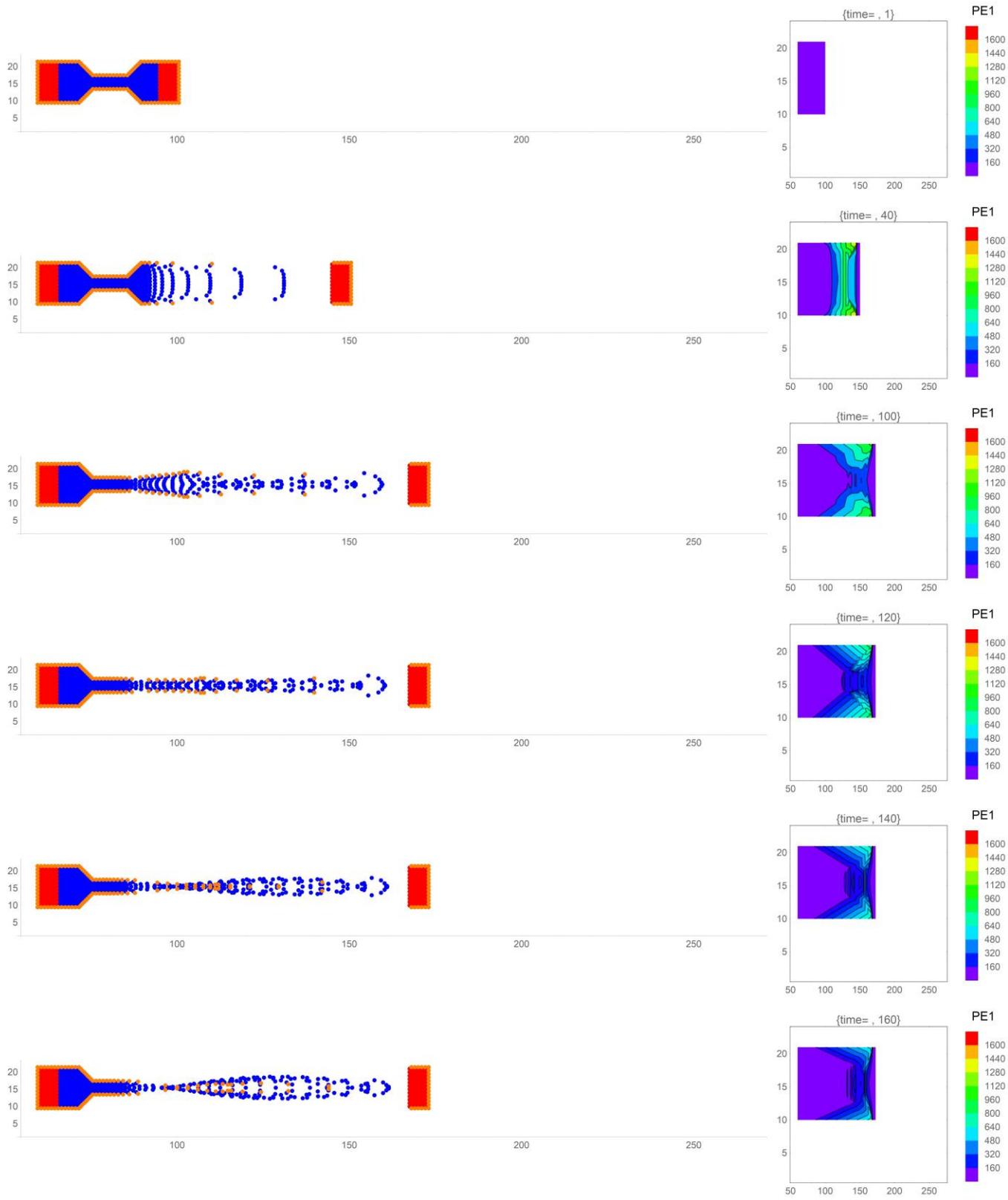

**Figure28** ASTM tensile test rectangular centred lattice with Poisson effect with coordination number five Configurations over different time (1, 40,100,120 and 160) with PE1 contour plot, of pseudoenergy.

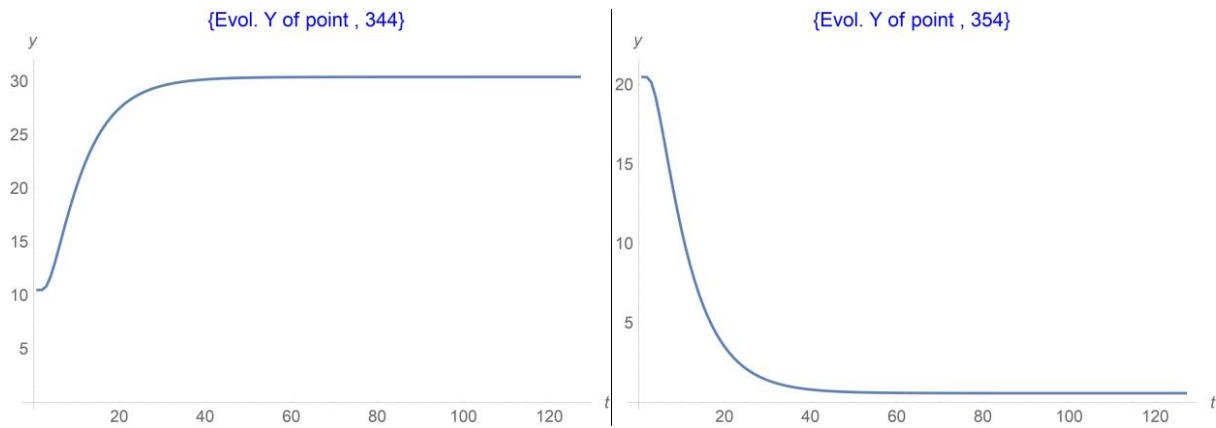

**Figure29** ASTM tensile test rectangular centred lattice with Poisson effect with coordination number five. Evolution Y coordinate of the point j=344 and 354 versus time; they always have the same X coordinate. Note: for scale reasons time is divided by 20, so last time step is 2500.

Looking at Figure29. We are considering two symmetric points so the evolution of x coordinate over the time is identical. The cross each other and this is, of course, impossible. This because the lateral contraction does not stop at y=15 (central line) but it goes further. This remembers as we are not connected with the physical phenomena and link with constitutive equations is a must.

The fracture test (see Figure30) is considered for the rectangular centered lattice with neighbor numbers to five. Distance fracture is 11 units and the speed was 0.6step/unit time. As can be seen, the fracture occurs at the top of the profile and not in the central area. Further studies, in progress, show that the fracture zone can be moved by varying working conditions. We can render the fracture more or less brittle changing the model parameters like neighbors' number, type of lattice, speed etc. As example in second gradient the same sample has a more brittle behavior; or if I use a speed of 2.5 step/unit time in the same condition I will get no followers on the right side of the fractured sample. The PE1 plot in this case is less significant because it does not take in account on the fictitious points; we are working on it to make this parameter meaningful also in fracture case.

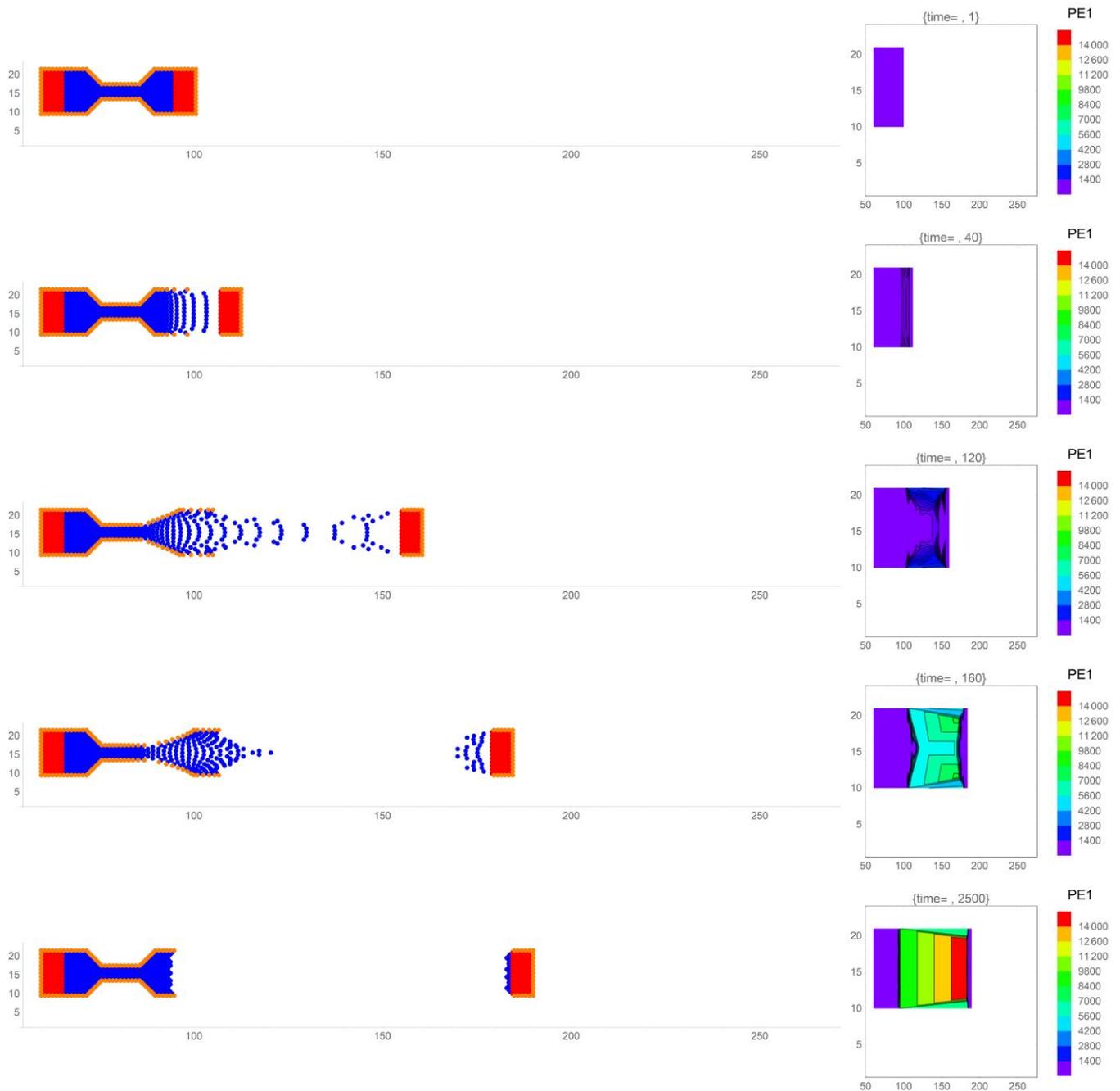

**Figure30**ASTM fracture test rectangular centred lattice, coordination number 5.

## 5. Conclusions

In this work we have presented many numerical simulations of discretized continuum deformation media modelled by a PBD method able to take into account a large variety of different behaviours.

Computational costs of the algorithm are lower with respect to FE analysis because we do not solve differential equations but only systems of algebraic equations; moreover working with a transformation operator between matrices the job can be parallelized between the GPU cores. Moreover, the proposed algorithm is intrinsically accounting for geometrically nonlinear deformations, which is a crucial theme in modern structural mechanics (see [53] for a general introduction, [54-56] for cases concerning systems presenting a geometry resembling the basic case of the proposed algorithm and [57-59], [31-33] for interesting results on large deformations of classical elastic models respectively in the static and dynamic case).

Based on preceding works we have changed lattice, rules and other parameters to demonstrate how very different results can be obtained, with the same strain condition. The same shape, in fact, shows different behaviour changing lattice, interaction rules of the follower and definition of the followers, to introduce first and second gradient theory. We have also used an ASTM shape whose results will be compared in a real tensile test experiment, but many questions are still open. The most important is the physical connection between constitutive equations of the material and the geometric constrains we impose on our model, to justify the utility of our tool. Practically if we want to compare our results with a tensile test experiment we have to connect constitutive parameters of the materials with our choice (lattice, etc...) in our tool. Finally, it would be interesting and challenging to investigate the homogenized limit of the proposed discrete model either using asymptotic methods [60-62] or (recalling that we introduced a discrete version of the deformation energy), using Gamma-convergence arguments (see [63] for a general introduction to the subject and [64-67] for applications to 1D and 2D elasticity). Work is in progress.